\documentclass{aastex701}
\usepackage{amsmath}
\usepackage{graphicx} 

\begin{document}


\title{Segmented-Polynomial-fitting Least Squares (SPLS): An optimized algorithm to find Earth twins}

\correspondingauthor{Fabo Feng}
\email{ffeng@sjtu.edu.cn}

\author{Shuyue Zheng}
\email{zhengshuyue@sjtu.edu.cn}
\affiliation{State Key Laboratory of Dark Matter Physics, Tsung-Dao Lee Institute \& School of Physics and Astronomy, Shanghai Jiao Tong University, Shanghai 201210, China}

\author{Fabo Feng}
\email{ffeng@sjtu.edu.cn}
\affiliation{State Key Laboratory of Dark Matter Physics, Tsung-Dao Lee Institute \& School of Physics and Astronomy, Shanghai Jiao Tong University, Shanghai 201210, China}

\author{Yicheng Rui}
\email{ruiyicheng@sjtu.edu.cn}
\affiliation{State Key Laboratory of Dark Matter Physics, Tsung-Dao Lee Institute \& School of Physics and Astronomy, Shanghai Jiao Tong University, Shanghai 201210, China}



\begin{abstract}

Detecting Earth twins remains challenging because their shallow, long-period transits are difficult to distinguish from background noise. Motivated by the challenge, we developed Segmented-Polynomial-fitting Least Squares (SPLS), a new algorithm that simultaneously fits planetary transits and background trends using a segmented double polynomial model. Prior to signal detection, the optimal polynomial order for the trend component is selected using Bayes factor-based model comparison. During the periodogram search, the Signal Detection Efficiency metric is used to assess signal significance. The algorithm is accelerated by a three-step approximation and nonlinear parameter sampling tailored to SPLS. We compare the performance of SPLS with traditional detrending-detection approaches across different orbital periods, signal-to-noise ratios (SNR), planet radii and stellar noise levels on an injection-recovery test. When detecting signals with periods between 10 and 480 days and SNRs below 9, SPLS achieves at least a 22.6\% higher true positive rate than other methods at the same 10\% false positive rate. Using the threshold determined from the Receiver Operating Characteristic curve analysis, our method also recovers the most true signals while yielding the fewest false positives among all injected samples, and reaches a 97\% recovery fraction in Kepler confirmed single-planet systems. The tests demonstrate that SPLS improves the detection of transiting planets, particularly for low-SNR, long-period signals. It offers the potential for finding Earth twins in future applications to data from Kepler, TESS, and upcoming PLATO and Earth 2.0 missions.

\end{abstract}
 


\section{Introduction} \label{sec:intro}

The transit method, which searches for brightness dips in stellar light curves induced by orbiting exoplanets, has emerged as an effective technique for exoplanet detection. It has accounted for the discovery of the majority of the over 6,000 confirmed exoplanets and produced more candidates awaiting follow-up conformation, largely enabled by space-based surveys such as Kepler \citep{2010Sci...327..977B}, K2 \citep{2014PASP..126..398H}, and TESS \citep{2015JATIS...1a4003R}. Nevertheless, no Earth twin has been successfully detected by any existing detection method to date. An Earth twin (Earth 2.0) is defined as an Earth-sized planet (0.8-1.25 $R_{\oplus}$) in the habitable zone of a solar-like star (F8V-K2V) (e.g., \citealt{2010Sci...327..977B}; \citealt{2013ApJ...766...81F}; \citealt{2022arXiv220606693G}), where the habitable zone refers to the orbital range within which a rocky planet's surface could support stable liquid water under appropriate atmospheric conditions (e.g., \citealt{1993Icar..101..108K}; \citealt{2013ApJ...765..131K}; \citealt{2017ApJ...837L...4R}). 

The pursuit of Earth twins holds profound significance, particularly in probing a core question of astrobiology and human curiosity: Are we the only intelligent life in the universe? Several Earth-like planets in the habitable zones of M-dwarfs have been discovered, such as Proxima b \citep{2016Natur.536..437A} and Trappist-1 e \citep{2017Natur.542..456G}), owing to their observational advantages such as the abundance of M-dwarfs and their closer-in habitable zones. However, the habitability of these planets may be threatened by tidal locking, which induces extreme temperature contrasts; intense stellar flares and XUV radiation, which can strip planetary atmospheres; and other factors (e.g., \citealt{2007AsBio...7...30T}; \citealt{2010AsBio..10..751S}). The search for habitable planets around K-dwarfs is also increasing, as K-dwarfs not only offer the observational advantages of low-luminosity stars but also exhibit lower stellar variability than M-dwarfs, suggesting a potentially better habitability (e.g., \citealt{2024ApJ...972...71L}; \citealt{2025A&A...694A..15B}). Although Earth twins around Sun-like stars are challenging to detect, we argue that solar-type stars remain the most promising systems in the search for biosignatures, as they provide the closest analogs to Earth's environmental conditions. The discovery of Earth twins would enable in-depth studies of their atmospheric properties, internal structures, and habitability. Establishing a statistically robust Earth-twin catalog with well-characterized properties would facilitate investigations into their occurrence rates and their correlations with stellar characteristics and galactic environments, thereby refining the theories of star-planet interactions and planetary formation and evolution. 

The transit detection of Earth twins is highly challenging, primarily due to their low signal-to-noise ratio (SNR) and an excessive number of false alarms and positives. In terms of transit signal strength, the geometric probability of transits for Earth twins is inherently low due to their long orbital periods. Even if a transit does occur, the small planetary radii result in a shallow transit depth of 84 ppm which is calculated based on Earth and the Sun. Additionally, the short observation time baseline limits the number of their detectable transits. Approximately three transits of an Earth twin can be detected in Kepler's four-year data which has the longest staring timeline among all missions to date. In terms of noise strength, intrinsic stellar variability which varies from star to star spans a wide range of timescales and amplitudes. Kepler results show that the stellar noise levels in quiet solar-type stars are unexpectedly high, exceeding the originally anticipated levels by about 50\% \citep{2011ApJS..197....6G}. Excess instrumental noise, such as readout noise and satellite operation, also reduces photometric precision (\citealt{2011ApJS..197....6G}; \citealt{2014PASP..126..948V}). Considering both the weak signal strength and relatively high noise levels, the SNR of Earth twins is very low. In addition to the challenge posed by the low SNR, the excess of false alarms and false positives at long periods is also a significant limiting factor (e.g., \citealt{2019AJ....157..143B}; \citealt{2025arXiv250103019I}). False alarms are caused by systematic effects and statistical detection methods, and false positives originate from various astronomical sources such as eclipsing binaries. The limitations of short staring timeline and instrumental noise are expected to be improved by upcoming PLATO \citep{2014ExA....38..249R} and Earth 2.0 \citep{2022arXiv220606693G} missions with enhanced photometric precision and extended observational baselines. Equally important, a sensitive algorithm capable of distinguishing Earth twins with low SNRs and long orbital periods from noise variability and reducing false positives and alarms is highly required.

The standard procedure for detecting planetary transits from photometric light curves involves first performing cotrending and detrending to remove correlated noise, and then searching for transit signals in the residuals with assumed white noise. Cotrending eliminates the common noise among nearby stars, primarily caused by instrumental and atmospheric effects, and detrending removes the unique noise of each star, primarily arising from intrinsic stellar variability. Detrending is mostly based on nonparametric or semiparametric techniques, such as the median or mean filter (e.g., \citealt{Moldovan_2010}; \citealt{Tal_Or_2013}; \citealt{Sinukoff_2016}), the Savitzky–Golay \citep{1964AnaCh..36.1627S} filter (e.g., \citealt{2011ApJS..197....6G}; \citealt{Howell_2014}), Gaussian process regression (e.g., \citealt{Aigrain_2015}; \citealt{Crossfield_2016}), wavelet transforms (e.g., \citealt{2002ApJ...575..493J}; \citealt{Carter_2009}; \citealt{Barclay_2015}) and singular spectrum analysis (e.g., \citealt{Boufleur_2017}; \citealt{fatheddin2024singularspectrumanalysisexoplanetary}). In the data with whitened noise, the widely used methods for searching periodic signals are the Box Least Squares (BLS; \citealt{2002A&A...391..369K}) and Transit Least Squares (TLS; \citealt{2019A&A...623A..39H}) periodograms, which phase-fold the light curve at different orbital periods, transit epochs and durations, and fit the phase-folded data with box-shaped and improved transit-shaped models respectively. Other methods, such as phase dispersion minimization \citep{1978ApJ...224..953S} and Lomb–Scargle periodograms \citep{1982ApJ...263..835S}, have been shown to be less effective than BLS \citep{2002A&A...391..369K}. In recent years, machine learning algorithms have been also increasingly applied to planetary transit detection (e.g., \citealt{2018AJ....155..147Z}; \citealt{Malik_2021}; \citealt{2024MNRAS.528.4053W}). 
 
The traditional step-by-step methods for processing noise and planetary signals outlined above have been successful in detecting high-SNR signals. However, these methods fail when the noise level is similar to the depth of the planetary signal and the number of transits is small (Earth twins). The main limitation is that, in blind searches, noise removal is based on the assumption of no planets. As a result, small planetary signals can be partially absorbed into the noise model, weakening the planetary signal in the residual light curve (\citealt{2015ApJ...806..215F}; \citealt{2019AJ....158..143H}; \citealt{2024AJ....167..284G}). Also detrending and cotrending may introduce distortions that mimic planetary signals. The limitations can be mitigated by simultaneously fitting the correlated noise and transit signals, which is utilized by \citet{2015ApJ...806..215F} and \citet{2024AJ....167..284G} to improve planetary detection in K2 light curves affected by high instrumental noise and in the light curves of active stars, respectively. Yet, the enhancement in the detection of these challenging signals is at the cost of higher computational expense than the standard step-by-step method, and the necessity of simultaneous fitting in practical applications has been partially questioned by \citet{2016A&A...585A..57K}. 

In this paper, we propose Segmented-Polynomial-fitting Least Squares (SPLS), a new algorithm that directly detects transits from cotrended light curves with background trends mainly from stellar variability. We use this approach to investigate the effectiveness of simultaneous fitting in detecting faint long-period signals, including Earth twins. We design a segmented double polynomial model to describe both transit signals and trends to handle the challenges posed by large time spans and complex noise. In our test, SPLS achieves a more sensitive performance compared to traditional approaches. The algorithm is also accelerated using an optimized three-step approximation, which fundamentally relies on combining the detection results of individual transits. The SPLS code is implemented in \texttt{Python} and available on GitHub\footnote{\url{https://github.com/Lu-dandandan/SPLS}}. The version used in this study can be found in Zenodo \citep{shuyue_zheng_2025_15411397}.

This paper is organized as follows. In Section \ref{sec: methods}, we introduce our method in detail. In Section \ref{section: performance}, we present an injection-recovery test in which the sensitivity of SPLS is compared with that of conventional detrending-detection methods, and we also apply SPLS to Kepler confirmed single-planet systems. Finally, Section \ref{sec: discussion} discusses the SPLS's advantages, limitations, scope of applicability, and prospects. Conclusions are drawn in Section \ref{sec: conclusion}.

\section{Methods}\label{sec: methods}

Our model of SPLS is applied to the light curves after cotrending and simultaneously fits transit signals star-specific background trends. Considering the variation in stellar photometry exhibits both long-scale and short-scale stochastic autocorrelated behaviors, which are challenging to model globally, we select subsets of data within several short windows centered on the sampled mid-transit times for fitting to reduce modeling complexity and focus more on signal detection. The optimal trend model is selected based on model comparison using the Bayes factor. To assess the significance of the transit signal in the periodogram, we employ log-likelihood difference and signal detection efficiency. Furthermore, the algorithm is accelerated by a three-step approximation and constraints on the range and resolution of parameter sampling. Based on the above considerations, SPLS has been developed.

\subsection{Segmented double polynomial model}

In order to improve the sensitivity of detecting small exoplanets, both background trends and transit signals are simultaneously modeled, which prevents transits from being distorted and weakened by prior denoising. Polynomials can describe a wide variety of transit shapes caused by different observational bands and impact parameters, and they have linear coefficients that facilitate solving, so we use two polynomials to model both the transits and the trends. A simplified segmented strategy, instead of global fitting, is also adopted due to the stochastic behavior of background variability. Specifically, we define a photometric data set $\{(t_i,\ f_i,\ \sigma_i)\}$, $i = 1,\ 2,\ \dots,\ n$, composed of three arrays of equal length that correspond to time, flux, and flux uncertainty, respectively. The window size is specified by the user before the search and only the data in the windows centered on the trial mid-transit times is fitted. Given the symmetrical nature of the transit signal, only even-order polynomials are utilized for modeling the signal. 

The periodic model intended to detect multiple transits of an exoplanet is 
\begin{equation}\label{eq: periodic transit}
\hat{f}_i =
\begin{cases} 
\sum\limits_k{A_k(t_i-t_{mj})^k}+\sum\limits_l{B_{lj}t_i^l} & t_i \in \left(t_{mj}-\frac{d}{2},\ t_{mj}+\frac{d}{2}\right) \\
\sum\limits_k{A_k(\frac{d}{2})^k}+\sum\limits_l{B_{lj}t_i^l} & t_i \in \left[t_{mj} - \frac{w}{2}, t_{mj} - \frac{d}{2}\right] \cup \left[t_{mj} + \frac{d}{2}, t_{mj} + \frac{w}{2}\right], 
\end{cases}
\end{equation}
where $w$ is the given window size required to be larger than sampled durations, $k\ (= 2,\ 4,\ \dots)$ denotes the degree of each term in the polynomial of the transit model without trends, and $l\ (= 0,\ 1,\ 2,\ \dots)$ denotes the degree of each term in the trend polynomial. The model has three nonlinear parameters, the first mid-transit time ($t_{m0}$), transit duration ($d$), and orbital period ($P$). A set of parameters $(P,\ d,\ t_{m0})$ corresponds to a set of transits and a set of windows. The mid-transit times of these transits are $t_{mj}$, where $j\ (= 0,\ 1,\ 2,\ ...)$ is the index number of the window. We assume that the transits share the same shape, but the background trends around them differ. Therefore, for different windows, the coefficients of the transit polynomial $A_k$ are the same and the coefficients of the trend polynomial $B_l$ are different. These coefficients are all linear parameters. $\sum\limits_kA_k(\frac{d}{2})^k$ is derived by taking the continuity of the transit boundary at $t_m \pm d$ into account.

\begin{figure}[ht!]
\includegraphics[width=\textwidth]{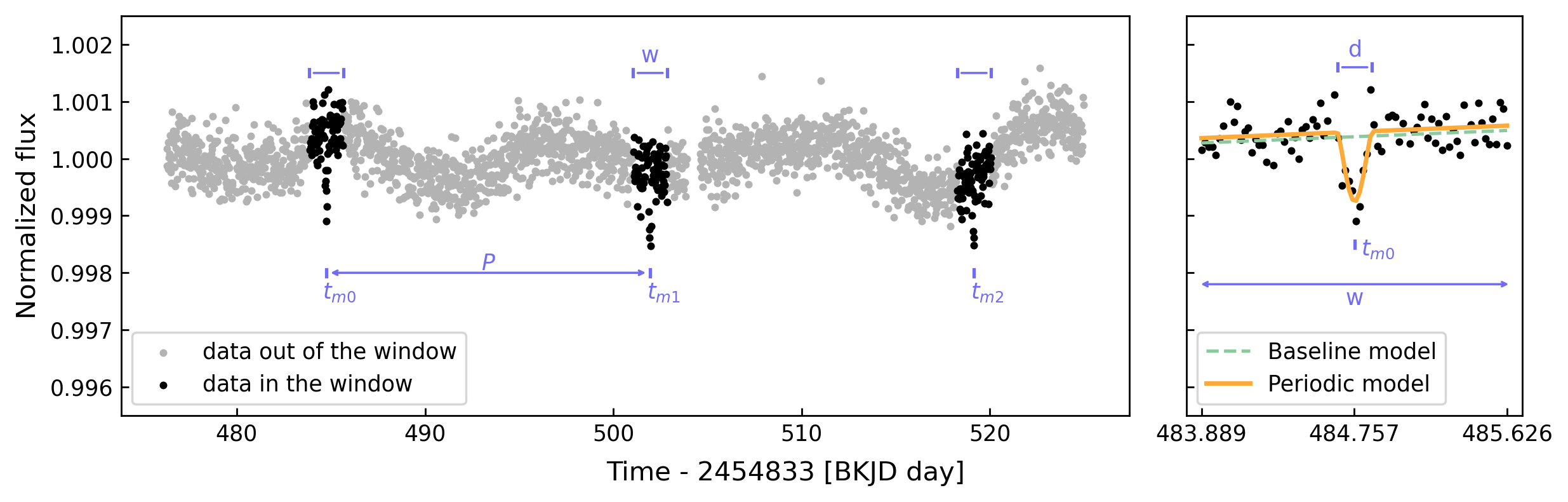} 
\caption{Illustration of the transit parameters with the Kepler light curve of Kepler-572. The light curve with a 29.4-min cadence in the left panel is after Kepler Pre-search Data Conditioning processing, removing points flagged as non-zero data quality and normalization in the time range of [475, 525] days in Quarter 5. For the best-fit parameters ($P$ = 17.205 d, $d$ = 3.6211 h, $t_{m0}$ = 2455008.058 d; \citealt{2016ApJ...822...86M}), black and gray points denote the data in and out of the 1.797-day window centered on the mid-transit times, respectively. The data points contained in the first transit window are shown in the right panel. The orange solid line and green dashed line are the fitted results using the periodic model and baseline model with a 4th-order polynomial for transits and a 1st-order polynomial for the trends, respectively.  
\label{fig:example_Kepler572}}
\end{figure}

A portion of the Kepler light curve of Kepler-572 is taken as an illustrative example in the following text. In the example, we use a fourth-order polynomial for transits and a first-order polynomial for the short-term trends which is determined by model comparison (see details in Section \ref{subsub: Model comparison of the stellar trend}). Here order means the degree of the highest-degree term. Figure \ref{fig:example_Kepler572} shows an intuitive illustration of the parameters of the periodic model. The example only presents the window positions for the best-fit parameters ($P$ = 17.205 d, $d$ = 3.6211 h, $t_{m0}$ = 2455008.058 d; \citealt{2016ApJ...822...86M}). In the blind search, the window positions vary with sampled nonlinear parameters. In addition, for each $(P,\ d,\ t_{m0})$, as shown in the right panel of Fig. \ref{fig:example_Kepler572}, the data in the windows is fitted with two models, periodic model and baseline model without parameter $A_k$, which will be compared to assess the significance of transit signal. 

A further constraint on the polynomial parameters of the transit component ($A_k$) is needed, as it is essential to ensure a concave transit shape and to avoid producing unrealistic false positives. It is obvious that the constraint is $A_2 > 0$ for a 2th-order transit polynomial. While the constraint on the fourth-order transit polynomial needs to be determined by calculating the first derivative and second derivative. ($A_2 > 0$ $\&$ $A_4 > 0$) and ($A_2 > 0$ $\&$ $\frac{A_2}{A_4}\leq -\frac{d^2}{2}$) are concluded to be allowed parameter space. If ($A_2 > 0$ $\&$ $\frac{A_2}{A_4}> -\frac{d^2}{2}$), the chi-square minimum will be searched along the line restricted by $\frac{A_2}{A_4}=-\frac{d^2}{2}$ in the parameter space. Higher-order polynomial models are not recommended for the transit component due to the increased derivation complexity of constraints and the corresponding higher computational expense.

\subsection{Statistic for periodogram}\label{subsec: Statistic for periodogram}
        
The significance of the transit signal at a given set of parameters $(P,\ d,\ t_{m0})$ is assessed utilizing the difference in the maximum log-likelihood between the periodic model and the baseline model in their respective linear parameter spaces firstly, which is defined as follows:
\begin{equation}\label{eq: dlnL}
\Delta\ln\mathcal{L}={\ln}\mathcal{L}_{\rm 1max}-{\ln}\mathcal{L}_{\rm 0max}, 
\end{equation}
where the subscripts 0 and 1 represent the baseline model and the periodic model, respectively. Assuming that the residuals between the model and the data points are Gaussian white noise, the logarithmic likelihood equation can be calculated by
\begin{equation}\label{eq: lnL}
\ln \mathcal{L}(P,\ d,\ t_{m0},\ \vec{\theta}_L)=-\frac{1}{2}\sum_i{\ln (2\pi\sigma_i^2)-\frac{1}{2}\sum_i\frac{(f_i-\hat{f}_i)^2}{\sigma_i^2}}=-\frac{1}{2}\sum_i{\ln(2\pi\sigma_i^2)}-\frac{1}{2}\chi^2, 
\end{equation}
where $\vec{\theta}_L$ represents all polynomial coefficients of the model, and $\chi^2=\sum_i\frac{(f_i-\hat{f}_i)^2}{\sigma_i^2}$. 
 
Specifically, given a parameter set of $(P,\ d,\ t_{m0})$, the polynomial coefficients  ($\vec{\theta}_L$) can be solved analytically by minimizing chi-square with the equation $\vec{\theta}_L=(X^{\rm T}WX)^{-1}X^{\rm T}WF$, where $X$, $W$, $F$ are the design matrix, weight matrix and the series of observed flux. Then $\ln \mathcal{L}_{\rm 1max}(P,\ d,\ t_{m0})$ and $\ln \mathcal{L}_{\rm 0max}(P,\ d,\ t_{m0})$ which are obtained by Eq. (\ref{eq: lnL}) with the solved coefficients are used to calculate the log-likelihood difference $\Delta\ln\mathcal{L}(P,\ d,\ t_{m0})$. Due to the lack of a single convex structure for $\Delta\ln\mathcal{L}$ in the three-dimensional nonlinear parameter space, it is necessary to sample $P$, $d$, and $t_{m0}$ to prevent missing the global optimal solution. Finally, the periodogram can be constructed by finding the two-dimensional maximum of $\Delta\ln\mathcal{L}$ at each period.

Furthermore, in order to apply a threshold and quantify how prominently the transit signal stands out from the background noise in each specific light curve, we convert the $\Delta\ln\mathcal{L}$ into the Signal Detection Efficiency (SDE; \citealt{2000ApJ...542..257A}), as BLS and TLS algorithms did. The difference of SDE between SPLS and them is that the quantity we used to calculate SDE is the normalized log-likelihood difference ($\widetilde{\Delta\ln\mathcal{L}}$), which is computed by shifting the raw $\Delta\ln\mathcal{L}(P)$ array to make its minimum value zero and then dividing the shifted array by its maximum value. The SDE is defined by 
\begin{equation}\label{eq: SDE}
{\rm SDE}(P) = \frac{\widetilde{\Delta\ln\mathcal{L}}(P)- <\widetilde{\Delta\ln\mathcal{L}}>}{\sigma(\widetilde{\Delta\ln\mathcal{L}})}, 
\end{equation}
where $<\widetilde{\Delta\ln\mathcal{L}}>$ and $\sigma(\widetilde{\Delta\ln\mathcal{L}})$ are the mean value and standard deviation of the normalized log-likelihood difference, respectively. SDE evaluates how many standard deviations the power in a periodogram is relative to the mean value. Obviously, the SDE of the potential transit signal with the highest peak in the periodogram is ${\rm \frac{1-<\widetilde{\Delta\ln\mathcal{L}>}}{\sigma(\widetilde{\Delta\ln\mathcal{L})}}}$. 

Similar to many periodograms, such as BLS, TLS, and transit comb filter (TCF) \citep{2019AJ....158...57C} periodograms, our periodogram also shows an increasing trend and scatter of the power as the period increases. The behavior is related to the smaller fractional transit duration at longer periods, the observational cadence, imperfect noise modeling, and other unknown factors. To reduce spurious long‐period signals, we follow the approach used by \citet{2014A&A...561A.138O} and TLS, which is removing the periodogram trend by subtracting the sliding median. The median‐filter window is set to more than ten times the period‐sampling oversampling parameter (see Section \ref{subsub: Sampling of nonlinear parameters}), so that peaks remain unaffected. Finally, SDE is recalculated from the detrended power array using Eq. (\ref{eq: SDE}).

\subsection{Three-step approximation}

The computational cost of strict SPLS is expensive, which is mainly caused by two reasons. On the one hand, the number $N$ of likelihood optimization ($2 \times N_P \times N_d \times N_{t_{m0}}$) of periodic model and baseline model depends on the size of the three-dimensional grid $(P,\ d,\ t_{m0})$. If we aim at searching for exoplanets with long orbital periods, the time span of the data points will be relatively long (over a thousand days), leading to the exceedingly high number of sampled periods $N_P$ (tens or hundreds of thousands). The sampling of $t_{m0}$ will also be performed with a finer resolution for accuracy. On the other hand, at each gird, phase folding, absence of binning, data indexing within windows, and matrix operation can also inevitably slow down the computation. Our tests show that analyzing a Kepler light curve with 17 quarters over tens of thousands of trial periods takes several days on a single CPU core. In short, the computational expense of this algorithm is substantial, which cannot be mitigated by only using a fraction of the data for fitting. The previous algorithms which simultaneously fit both noise and signal also suffer from the limitation of slow speed and were subsequently simplified  (\citealt{2015ApJ...806..215F}; \citealt{2024AJ....167..284G}). Based on their simplified approach, we implemented a similar, but enhanced, three-step approximation to accelerate our algorithm.

The three-step approximation consists of linear search, period search, and global fitting. The main idea is to relax the assumption that all transits, for a given set of nonlinear parameters, have the same shape. $A_k$ can differ for transits occurring at different transit epochs. Therefore, the original rigorous fitting approach can be dismantled and restructured efficiently, as shown in the following equation
\begin{equation}
\ln \mathcal{L}(P,\ d,\ t_{m0},\ \vec{\theta}_L) = \sum_j\ln\mathcal{L}(P,\ d,\ t_{m0},\ A_k,\ B_{lj}) \approx \sum_j\ln\mathcal{L}(P,\ d,\ t_{m0},\ A_{kj},\ B_{lj}).
\end{equation}
 
\begin{figure}[ht!]
\includegraphics[width=\textwidth]{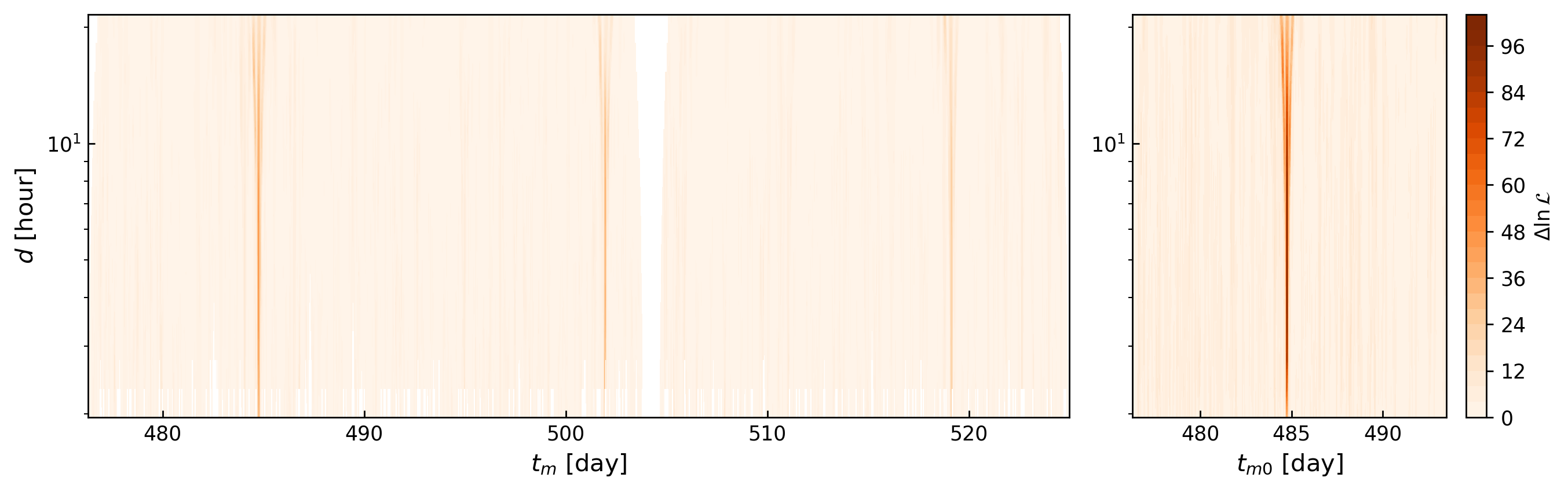} 
\caption{$\Delta\ln\mathcal{L}$ distribution in the linear search and periodic search of the example light curve of Fig. \ref{fig:example_Kepler572}. The distribution of the log-likelihood difference in ($d$, $t_{m}$) grids of linear search is plotted in the left panel. The same polynomial order and window size are set as Fig. \ref{fig:example_Kepler572}. 15 durations were uniformly sampled in logarithmic space and also 28,708 mid-transit times were uniformly sampled. The number, 28,708, is determined by the set period minimum of 1.797 days (See the principle of parameter setting in Section \ref{subsub: Sampling of nonlinear parameters}). The right panel shows the new distribution of $\Delta \ln\mathcal{L}(d,\ t_{m0})$ by reshaping the $\Delta\ln\mathcal{L}(d,\ t_m)$ of the left panel and reducing dimension at the best period (17.205 d).
\label{fig:gird2D}}
\end{figure}

In the linear search, we generate $\Delta\ln\mathcal{L}$ distribution on a two-dimensional grid in transit durations and mid-transit times which cover the time span of the entire light curve uniformly. For each pair of $t_m$ and $d$, likelihood maximization is performed to obtain $\ln\mathcal{L}_{1{\rm max}}$ for the single transit model and $\ln\mathcal{L}_{0{\rm max}}$ for the baseline model by using the data within a window centered on $t_m$. The single transit model here refers to the model without the period information and the index $j$ from Eq. (\ref{eq: periodic transit}). Then $\Delta\ln\mathcal{L}(d,\ t_{m})$ is calculated by Eq. (\ref{eq: dlnL}). The left panel of Fig. \ref{fig:gird2D} shows the distribution of log-likelihood difference for the light curve in the left panel of Fig. \ref{fig:example_Kepler572}. Three transits with large $\Delta\ln\mathcal{L}$ can be seen clearly. To make the distribution more continuous and easier to visualize here, no constraint on $A_k$ was applied in the example.

In the periodic search, we set a series of trial periods that are integer multiples of the constant sampling interval of $t_m$, and sum $\Delta\ln\mathcal{L}$ at different transits for one ($P$, $d$, $t_{m0}$). Specifically, for each period, the two-dimensional matrix $\Delta\ln\mathcal{L}(d,\ t_m)$ of the first step is reshaped into a three-dimensional matrix $\Delta\ln\mathcal{L}(d,\ t_{m0},\ t_{mj}),\ j=1,\ 2,\ ...$, which is then reduced to a new 2D matrix $\Delta\ln\mathcal{L}(d,\ t_{m0})$ by summing over the third dimension $t_{mj}$. The optimal values of $t_{m0}$ and $d$ with the highest log-likelihood difference for that period can be determined. The right panel of Fig. \ref{fig:gird2D} presents an example for the best orbital period of the planet.

The third step is global fitting. If we directly construct the periodogram based on the results of the second step, several issues will arise. The main issue is that the degrees of freedom of the approximate model are relatively large and depend on the number of transits. More transits at smaller periods tend to increase the log-likelihood difference. Therefore, in this step, the more rigorous transit model, the periodic model (\ref{eq: periodic transit}), and the baseline model are applied to fit the light curve for each trial period and its corresponding best $t_{m0}$ and $d$ of the second step. Eventually, the periodogram is constructed with the calculated log-likelihood difference and SDE for a range of periods.

Apart from the model, segmented data, and statistics, our three-step approximation also has some other differences compared to the previous simplified algorithms. One is that the likelihood formula with depth correction, as implemented in Nuance (Eq. (8) in their paper; \citealt{2024AJ....167..284G}), is not used in our approach. Our internal tests showed that the correction had a negligible impact on our statistical measure. Besides, instead of using interpolation or nearest neighbors, we directly reshape the 2D ($d,\ t_m$) grid in the periodic search. Furthermore, the fitting is applied to partial data only, and other places with transits will not be mistaken as no signal by the single transit model.

\subsection{Sampling of nonlinear parameters}\label{subsub: Sampling of nonlinear parameters}

We adopt the period sampling method of \citet{2014A&A...561A.138O} and \citet{2019A&A...623A..39H} which is sensitive and effective to reduce computational cost. They deduced that the optimal sampling interval for frequency is frequency-dependent rather than a constant: 
\begin{equation}\label{eq: dfre}
\begin{split}
\mathrm{d}f &= Af^{2/3} \\
A  &\equiv \frac{(2\pi)^{2/3}}{\pi}\frac{R_{\rm s}}{(GM_{\rm s})^{1/3}}\frac{1}{S\times OS}, 
\end{split}
\end{equation}
and the solution is employed as the optimal frequency sampling: 
\begin{equation}\label{eq: sampling of periods}
\begin{split}
f(x) &= (\frac{A}{3}x+C)^3   \ \ \  x = 1, 2, 3, \dots\\
C  &\equiv f^{1/3}_{\rm min}-\frac{A}{3}, 
\end{split}
\end{equation}
where $f$ denotes orbital frequency, $R_{\rm{s}}$ and $M_{\rm{s}}$ are the radius and mass of the host star respectively, $G$ is the gravitational constant, $S$ is the time span of the light curve, and $OS$ is the oversampling parameter with a value between 2 and 5 to adjust the resolution near the optimal peak. The maximum sampled period ($P_{\rm {max}}$) is set to $S/3$ (or $S/2$). However these periods cannot be used directly in our algorithm due to our design that the sampled periods are generated by folding $t_m$. Therefore, each sampled period from Eq. (\ref{eq: sampling of periods}) must be approximated to the nearest integer multiple of $\Delta t_m$ to be compatible with our algorithm.

\begin{figure}[ht!]
\centering
\includegraphics[width=0.9\linewidth]{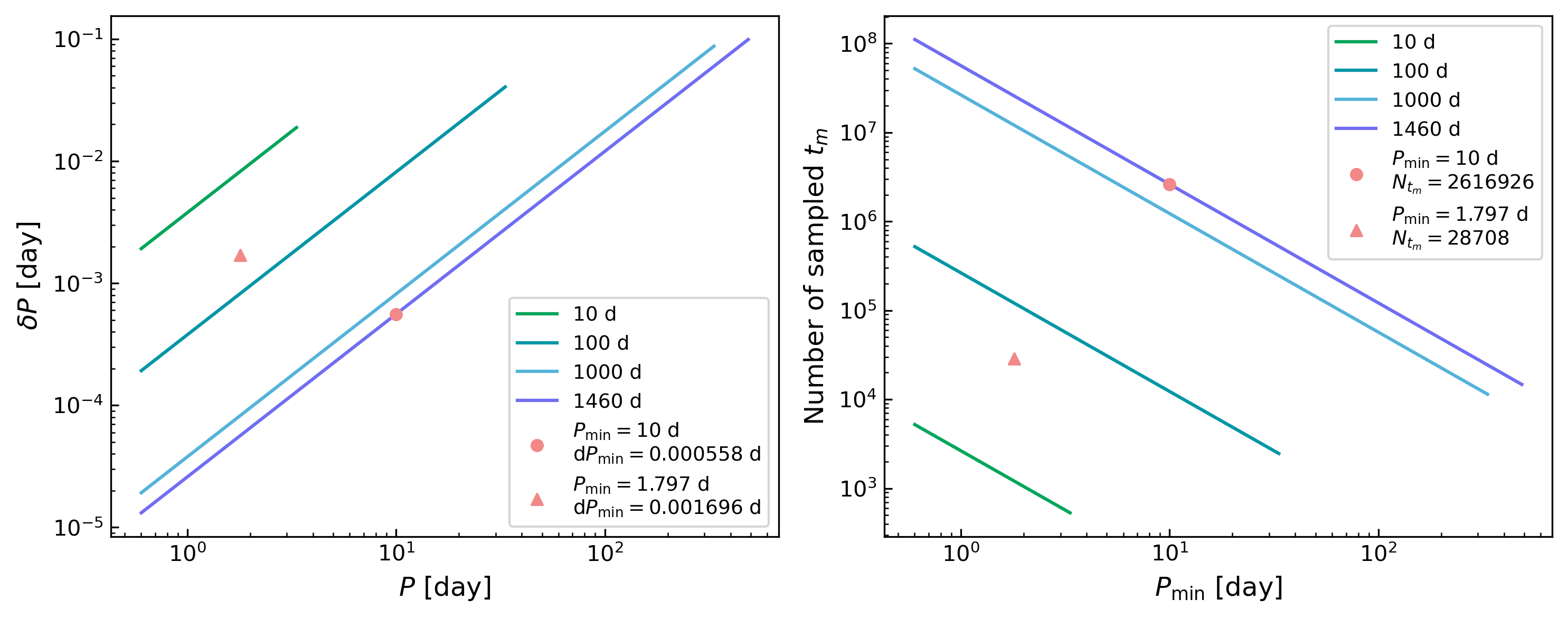} 
\caption{Period resolution ($\delta P$) and the number of sampled mid-transit times as functions of trial periods. The lines are expressed in Eq. (\ref{eq: dP}) (left) and Eq. (\ref{eq: n_tm}) (right) with four time spans of light curves and all are linear in logarithmic space. The pink circle represents the 10-day minimum period for the full Kepler light curve. The pink triangle represents the default parameter setting for the example light curve of Kepler-572 ($S=48.7\ {\rm d}$). 
\label{fig: dP}}
\end{figure}

Since trial periods are produced by folding $t_m$, in other words, $\Delta t_m$ determines the resolution of the periods ($\Delta P$). This raises a question: how fine should the $t_m$ grid be to achieve sufficient period resolution? $\Delta t_m$ should not be larger than the minimum interval of periods ($\Delta P_{\rm {min}}$). Here we only consider $\Delta t_m$ = $\Delta P_{\rm {min}}$ to minimize the number of sampling as much as possible. To find $\Delta P_{\rm {min}}$, we convert Eq. (\ref{eq: dfre}) into the period incremental Eq. (\ref{eq: dP}), which is shown in the left panel of Fig. \ref{fig: dP}. From the figure, $\Delta P_{\rm {min}}$ is found at the minimum of trial periods ($P_{\rm {min}}$) for a given light curve. Therefore, once the minimum of sampled periods is set, the mid-transit time interval is determined: $\Delta t_m = \Delta P_{\rm {min}} = AP^{4/3}_{\rm {min}}$.
\begin{equation}\label{eq: dP}
\delta P = AP^{4/3}
\end{equation}
\begin{equation}\label{eq: n_tm}
N_{t_m}=\frac{S}{\Delta t_m}=\frac{S}{\mathrm{d}P_{\rm {min}}}=\frac{S}{A}P^{-4/3}_{\rm {min}}
\end{equation}

Here, we expect to determine a reasonable minimum period ($P_{\rm min}$). The default minimum period is equal to the window size $w$ to prevent the repeated use of data for one ($P,\ d,\ t_{m0}$), while sometimes a larger minimum of trial periods is a more appropriate choice. It is because $P_{\rm min}$ directly determines the number of sampled mid-transit times ($N_{t_m}$) by Eq. (\ref{eq: n_tm}) and subsequently affects the computational expense. From the equation, $N_{t_m}$ is related to the time span and the minimum period when the stellar mass, radius, and oversampling parameters are fixed. From the right panel of Fig. \ref{fig: dP}, $N_{t_m}$ increases with the extension of the time span for the same minimum period, and decreases exponentially as the minimum period increases for the same time span, indicating that users can set $P_{\rm min}$ by balancing search time and their detection goal. The pink triangle marks the default parameter setting ($P_{\rm min}=w=1.797$ d) for the example light curve presented earlier whose time span is relatively small ($S=48.7\ {\rm d}$). If we aim to search for planets with orbital periods of several hundred days, a slightly longer minimum period is recommended for tractable running time. For a Kepler light curve observed for 17 quarters whose time span is about 1460 days, we adopted a minimum period of 10 days (the pink circle). 

\begin{figure}[ht!]
\centering
\includegraphics[width=0.9\linewidth]{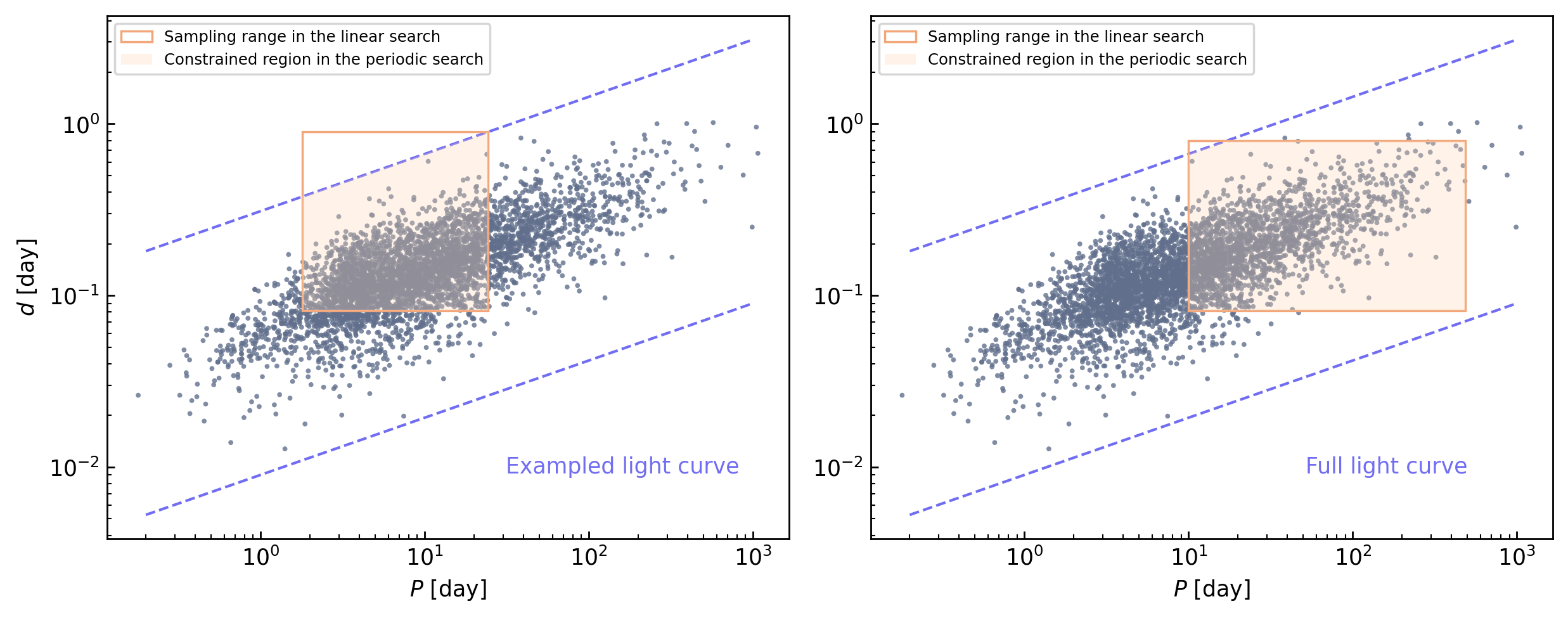} 
\caption{The relation of orbital period and transit duration of confirmed transiting exoplanets from NASA Exoplanet Archive. EPIC 248847494 b with a 3650-day period is not plotted in the figure because of the single transit. Orange region is the sampled range of period and duration for the exampled light curve of Kepler-572 (left; default sampling) and a full long-cadence Kepler light curve observed for 17 quarters with 0.8-day $d_{\rm max}$ (right). The region enclosed by the solid orange line represents the sampling in the linear search, and the shaded area represents the further constrained sampling in the periodic search. \label{fig: confirmed_planets}}
\end{figure}

The default sampling of duration in SPLS is based on the distribution of transit duration and orbital period for the confirmed transiting exoplanets from NASA Exoplanet Archive\footnote{\url{https://exoplanetarchive.ipac.caltech.edu}} which is shown in Fig. \ref{fig: confirmed_planets}. The purple lines follow the relation $d = EP^{1/3}$ which is derived from Kepler's third law and the assumption of circular orbits. $E$ is a constant that is mainly related to the parameters of the star. We adjusted $E$ so that all the confirmed transiting planets lie between the two dashed lines. The minimum of trial duration $d_{\rm min}$ depends on the cadence ($\Delta t$) of the data by $(k_{\rm max}/2+2) \times {\Delta t}$ where $k_{\rm max}$ is the polynomial order for the transit component\footnote{The extra factor of 2 is added for the stability of matrix operation.}, which makes sure enough data points within transit duration for calculating $A_k$. We show the default duration sampling with an inputted value $d_{\rm max}=21.564$ h for the exampled light curve of Kepler-572 in the left panel. Then durations are sampled uniformly in the logarithmic space. The region enclosed by the orange border represents the sampling range in the linear search. To accelerate the computation, in the periodic search, we further constrain the sampling region within the area between the purple dashed lines, which is shown as the orange-shaded region. The summary of all the default parameter settings is provided in Table \ref{tab:nonlinear parameters}. Users are allowed to define their own sampling method according to their requirements, as long as the sampling complies with the loosest constraints (the third column of Table \ref{tab:nonlinear parameters}). 

\begin{table} 
\renewcommand\arraystretch{1.1}
\begin{center}
\caption{{Sampling of nonlinear parameters and the setting of window size.}
\label{tab:nonlinear parameters}}
\begin{tabular}{lcc}
\hline
Parameters            & Default values                                                                        & Necessary constraints on user-defined values  \\
\hline
$P_{\rm max}$         & $S$/2 or $S$/3                                                                        &  $\leq S$/2                                    \\
$d_{\rm min}^{(1)}$   & $(k_{\rm max}/2+2)\ \times \ \Delta t$                                                         & $\geq (k_{\rm max}/2+2)\ \times \ \Delta t$           \\
\cline{3-3}
$d_{\rm max}^{(1)}$   & input                                                    &                                          \\ 
$w$                   & input                                                                   &{$d_{\rm max}\ < w\ \leq P_{\rm min}$}   \\
$P_{\rm min}$         & $w$                                                                                   &                                         \\
\cline{3-3}
$\Delta P$            &  Eq. (\ref{eq: dP})                                    & -                                       \\
$\Delta d^{(1)}$      &  Uniform in logarithmic space                                                         & -                                      \\
$t_{m\ {\rm min}}$    & $t_{\rm min}$                                                                         & $t_{\rm min}$                          \\
$t_{m\ {\rm max}}$    & $t_{\rm max}$                                                                         & $t_{\rm max}$                        \\
$\Delta t_m$          & Uniform, min\{$d_{\rm min}^{(1)}/OS_{t_m}$, $AP^{4/3}_{\rm {min}}$\}$^a$                    & Uniform, min\{$d_{\rm min}^{(1)}/OS_{t_m}$, $\Delta P_{\rm min}$\}$^a$\\
$d^{(2)}(P)$          &  $d^{(1)}\cap$ the area between two dashed lines in Fig. \ref{fig: confirmed_planets} & Optional    \\
$\Delta t_{m0}^{(2)}$ & An integer multiple of $\Delta t_m$ closest to $d^{(2)}_{\rm min}(P)/OS_{t_m}$$^b$    & An integer multiple of $\Delta t_m$ closest to $d^{(2)}_{\rm min}(P)/OS_{t_m}$$^b$ \\

\hline
\end{tabular}
\end{center}
$^{(1)}$ The sampling in the linear search.  

$^{(2)}$ The sampling in the periodic search.    

$^a$ In practice, $\Delta t_m$ is the minimum value between $d_{\rm min}/OS_{t_m}$ and $\Delta P_{\rm min}$, where $OS_{t_m}$ (1$\sim$10) is the oversampling parameter for $t_m$. Usually, $\Delta P_{\rm min}$ is smaller, so in the text, $\Delta t_m = \Delta P_{min}$ is shown only.

$^b$ If $\Delta P_{\rm min}$ is smaller compared to $d_{\rm min}^{(2)}/OS_{t_m}$, after folding at each $P$, the resolution of $t_{m0}$ is expanded to $d_{\rm min}^{(2)}(P)/OS_{t_m}$ which is approximated to the closest integer multiple of $\Delta t_m$, accelerating the periodic search of SPLS.
\end{table}

\subsection{Trend model selection}\label{subsub: Model comparison of the stellar trend}
\begin{figure}[ht!]
\centering
\includegraphics[width=0.5\linewidth]{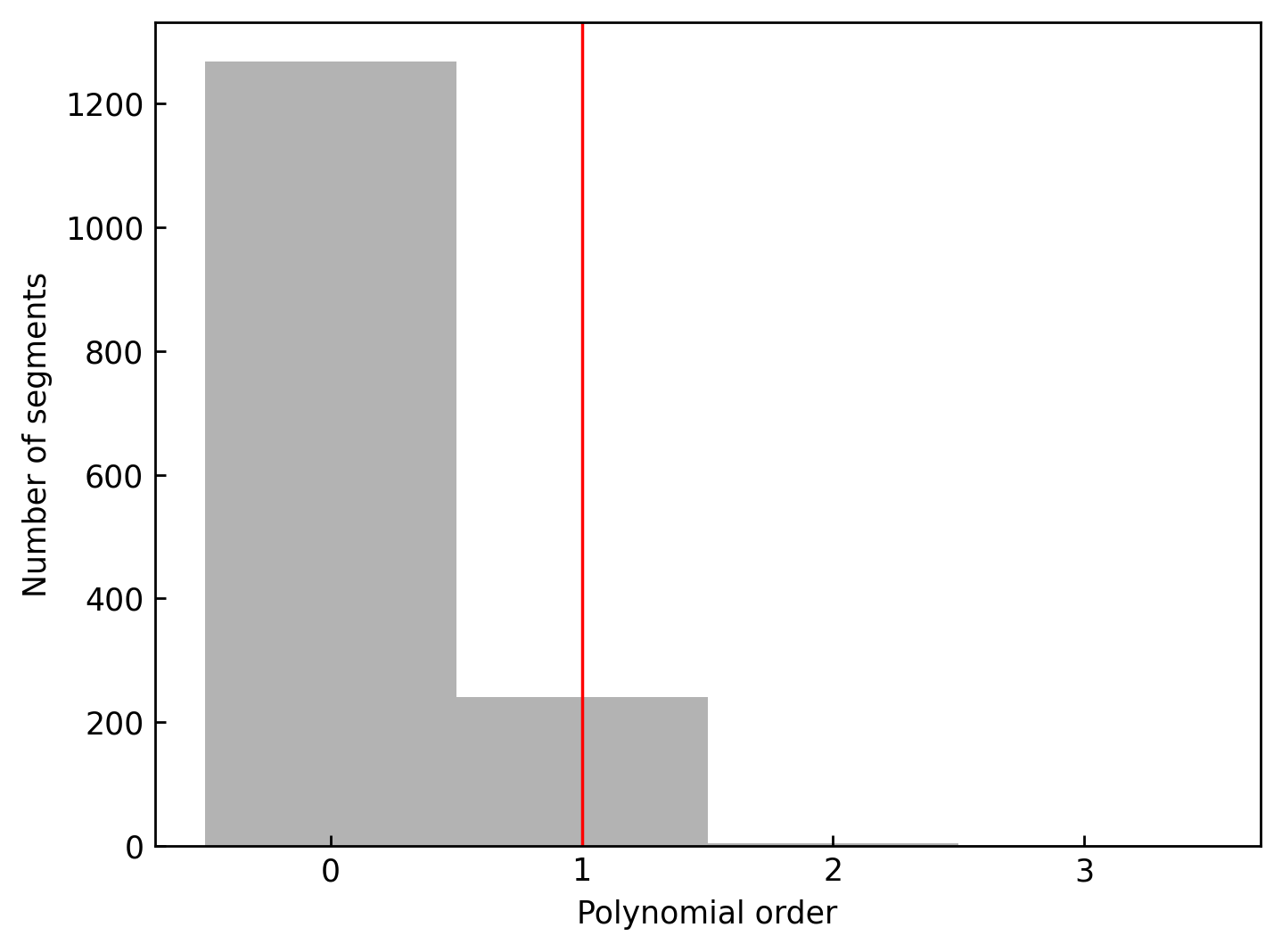} 
\caption{Frequency histogram of best trend orders of all segments of the example light curve of Kepler-572. The red line shows the 90th percentile of the distribution whose corresponding trend order, 1-st order, is chosen as the optimal trend order for the target finally.
\label{fig: N_orders}}
\end{figure}

Since noise levels vary between stars, the optimal polynomial order for the background trends of each light curve is determined through model comparison. The BIC-converted Bayes factor is adopted as the metric, as \citet{2016MNRAS.461.2440F} did in the radial velocity data analysis. Similar to the linear search, we segment the data based on the window size of SPLS. The mid-time of each segment is uniformly sampled across the entire time span, and dense sampling can lead to the overlap of data segments. For each segment, we fit the data with a polynomial model from 0-th order to 3-th order without modeling transits and calculate the corresponding logarithmic Bayes factors ($\ln{\rm BF}_{ij}$). The maximum order can be set to larger than 3-th order by users. Owing to the difficulty of calculating the posterior, the Bayesian information criterion (BIC) is utilized to estimate the Bayes factor \citep{Kass01061995}. The formulae are
\begin{equation}
{\rm BIC} = -2 \ln \mathcal{L}_{\rm max}+m \ln N, 
\end{equation}
\begin{equation}
\rm \ln BF_{ij} \approx \frac{BIC_j-BIC_i}{2}, 
\end{equation}
where the subscripts $i$, $j$ = 0, 1, 2, 3 represent the different order of the trend model, $\mathcal{L}_{\rm max}$ is the maximum value of likelihood function, $m$ is the number of free parameters, and $N$ is the number of data points. $\ln{\rm BF}_{ij}$ larger than 5 means $i$-th order model is preferred by data to $j$-th order model with a strong evidence (\citealt{Kass01061995}; \citealt{2016MNRAS.461.2440F}). Models with adjacent complexities are compared using this criterion until the best model is identified. Table \ref{tab: lnBFi0} displays partial $\ln{\rm BF}_{i0}$ results of Kepler-572. The optimal trend model for each segment is highlighted in boldface. The optimal trend orders of all segments of the example light curve are plotted in the frequency histogram (Fig. \ref{fig: N_orders}). We select the 90th percentile as the optimal order for the entire light curve. The 90th percentile is a subjective choice. If the transit occurs in a region with more complex noise, a larger-order trend model may better capture the characteristics of the noise.

\begin{table} 
\renewcommand\arraystretch{1.1}
\begin{center}
\caption{{The logarithmic Bayes factors $\ln{\rm BF}_{i0}$ of trend polynomial models for partial segments of Kepler-572. Totally 1514 segments are sampled. For each segment, the best trend model is highlighted in boldface.} 
\label{tab: lnBFi0}}
\begin{tabular}{cccccc}
\hline
Trend order&$\ln{\rm BF}_{i0}$ for Segment 0&...&$\ln{\rm BF}_{i0}$ for Segment 1505&$\ln{\rm BF}_{i0}$ for Segment 1506 & ... \\
\hline
0          & \textbf{0.0000}           &...& 0.0000                       & 0.0000                        & ... \\
1          & 0.3328                    &...& \textbf{10.9657}             & 14.3717                       & ... \\
2          & -1.4094                   &...& 15.4549                      & \textbf{25.1099}              & ... \\
3          & -3.2281                   &...& 13.7749                      & 28.1863                       & ... \\
\hline
\end{tabular}
\end{center}
\end{table}

\section{Performance}\label{section: performance}

To evaluate the performance of our algorithm and its feasibility of detecting Earth twins, we conducted an injection-recovery test on Kepler light curves and compared the results of SPLS with common detrending-detection methods, biweight\footnote{The implementation in \texttt{W$\overline{o}$tan} Python package is used.}+BLS\footnote{The implementation of \texttt{BoxLeastSquares} of \texttt{Astropy} is used.} and biweight+TLS in Section \ref{subsection: injection-recovery test}. ``Biweight" refers to a time-windowed slider with an iterative robust location estimator based on Tukey’s biweight and is identified as the optimal detrending tool by \citet{2019AJ....158..143H}. BLS and TLS algorithms are then applied to detect the transits after biweight detrending. In addition, we applied our algorithm to Kepler confirmed single-planet systems (see Section \ref{subsection: application to Kepler Single-Planet Systems}) to further evaluate its performance on real data.

\subsection{Injection-recovery test}\label{subsection: injection-recovery test}

\subsubsection{Test data and preprocessing}\label{subsubsection: Test data and preprocessing}

Since our method simultaneously models both background trends and transiting signals, accurately simulating the noise in the test data is crucial. To ensure realism, we selected real Kepler light curves as test samples for the injection-recovery test, avoiding any subjectivity introduced by artificially simulated noise. We performed a bulk download of all 207,617 long-cadence (29.4-min) Kepler DR25 light curves from MAST and used the flux that has been corrected for instrumental effects by the Kepler Pre-search Data Conditioning (PDC) process. All the Kepler data used in this paper can be found in \dataset[10.17909/T9488N]{http://dx.doi.org/10.17909/T9488N}. Eclipsing binaries identified in the Kepler mission \citep{2016AJ....151...68K}, Kepler Objects of Interest (KOIs) \citep{2018ApJS..235...38T}, and the host stars of confirmed planets\footnote{\url{https://exoplanetarchive.ipac.caltech.edu/cgi-bin/TblView/nph-tblView?app=ExoTbls&config=kep_conf_names}} are excluded to ensure that the selected targets are as free of planets as possible. The light curves may occasionally contain small planetary transits that have not been detected by any algorithm, but we approximate that the potential transits will not affect the comparison between algorithms \citep{2024AJ....167..284G}. We also only considered the targets that were observed for the full 17 quarters (Q1–Q17) to maximize data availability over the four-year observation. Data from Quarter 0 was excluded due to its poor quality. A total of 106,658 light curves passed these selection criteria. 

To further ensure that our samples were suitable for planet injection, we selected targets with complete stellar properties (mass, radius, effective temperature, surface gravity, and metallicity) by cross-matching with the Gaia–Kepler Stellar Properties Catalog \citep{2020AJ....159..280B}. We roughly excluded evolved stars by applying  selection criteria of $R_{\rm{s}}$ $<$ 1.5 R$_{\odot}$ and $\log g$ $>$ 4.1. The reasons are that evolved stars typically have larger radii and higher background noise, making planet signals harder to detect, and since we focused on weak long-period planets such as Earth twins, we prioritized near solar-type stars for their stability and suitability for such studies. After these cuts, 62,352 targets remained, and 3,000 of them were randomly selected to build the light curve library for the injection–recovery test.

We created 10,000 light curves of transiting single-planet systems by injecting planetary signals into randomly selected light curves from the light curve library. The injected orbital periods were drawn log-uniformly between 10 and 480 days. Periods shorter than 10 days were not considered due to the higher computational cost of our algorithm (see Fig. \ref{fig: dP}), while the upper limit of 480 days ensured at least three transit events for each injected planet. The injected transit durations were then calculated using Kepler’s third law, under the assumptions of circular orbits and zero impact parameter. We sampled the transit SNRs or planetary radii ($R_{\rm p}$) according to the equation:
\begin{equation} 
SNR = (\frac{R_{\rm p}}{R_{\rm s}})^2\frac{\sqrt{N}}{\sigma}, 
\end{equation}
where $\sigma$ is the noise level and $N$ is the number of transits. Among the 10,000 samples, 3,000 planets were injected with radii uniformly distributed from 0.5 to 12 R$_{\oplus}$, 3,000 with SNRs uniformly drawn from 4 to 20, and 4,000 with SNRs drawn from a $\beta$-distribution ($\alpha$ = 2, $\beta$ = 5) over the same range. This sampling design enables a comprehensive evaluation of the algorithm across a wide range of planetary radii and SNRs, with finer sampling near the detection limits for better sensitivity characterization. Additionally, the noise level $\sigma$ of the light curve is characterized by Combined Differential Photometric Precision (CDPP) which is calculated using the \texttt{Lightkurve} python package \citep{2018ascl.soft12013L}. The two input parameters of \texttt{LightCurve.estimate\_cdpp}, ``transit\_duration" and ``savgol\_window", were respectively set to half and three times the injected duration. We also set the epoch of the first transit in the data randomly, and used a non-linear limb-darkening model to generate the light curve, where the limb-darkening coefficients were obtained according to \citealt{claret2000}. 
 
We conducted a straightforward preprocessing of the light curves. Each quarter's data was normalized by dividing by its median value. We then discarded PDCSAP\_FLUX data with NaN values or nonzero quality flags, as these flags indicate various systematic issues. This cleaning helps reduce outliers to some extent, but a few outliers still remain. Due to their potential similarity to transits, the remaining outliers were not further processed.

\subsubsection{Settings of three methods}\label{subsubsection: settings of three methods}

In the test, both light curves with and without transits are searched using the three methods: SPLS, biweight+BLS, and biweight+TLS. The window size of the biweight filter is set to three times the injected transit duration which is the optimal width \citep{2019AJ....158..143H}. For convenience, the window size of SPLS is set to the same value as that of the biweight filter. The trial periods for BLS, TLS, and SPLS are sampled based on Eq. (\ref{eq: sampling of periods}) with $M_{\rm s}=1M_{\odot}$, $R_{\rm s}=1R_{\odot}$, $OS=2$, $P_{\rm min}=10\ {\rm d}$ and $P_{\rm max}=S/3$, generating about 39,029 trial periods. The only slight difference is that each sampled period in SPLS is further approximated to its nearest value which is an integer multiple of the interval of $t_m$ in the linear search, as we mentioned earlier. This difference is negligible in the comparison. The transit duration sampling in TLS is performed using its built-in method. BLS and SPLS sample 15 durations uniformly in logarithmic space between the default $d_{\rm min}$ listed in Table \ref{tab:nonlinear parameters} and $d_{\rm max}$ which is set to half of the window size. The right panel of Fig. \ref{fig: confirmed_planets} shows an example of the sampling ranges for periods and durations used by SPLS. Furthermore, the minimum search depth for TLS is set to 1 ppm, the number of bins per duration for BLS is fixed at 15, and all remaining SPLS parameters are maintained at their default values, as specified in Table \ref{tab:nonlinear parameters}.

\begin{figure}[ht!]
\centering
\includegraphics[width=\linewidth]
{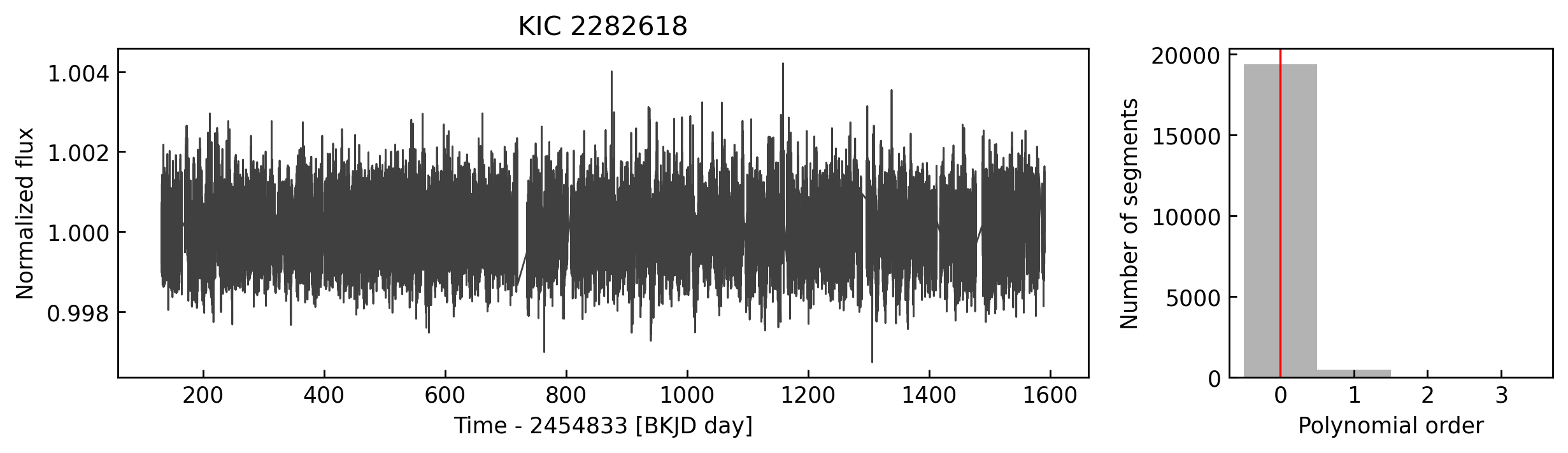}
\includegraphics[width=\linewidth]
{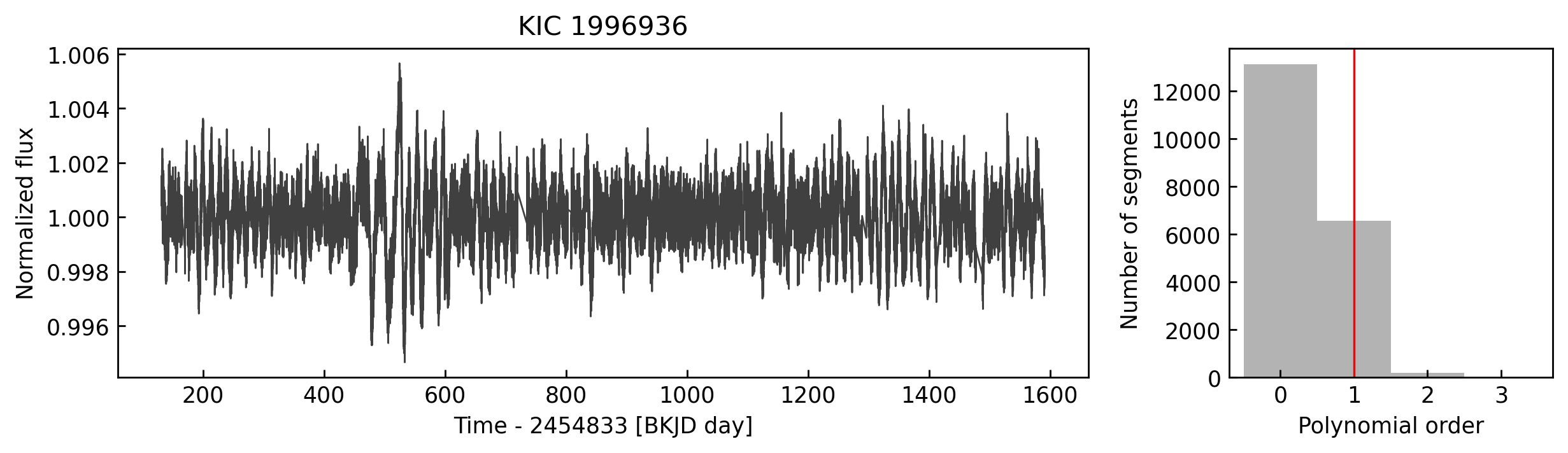}
\includegraphics[width=\linewidth]
{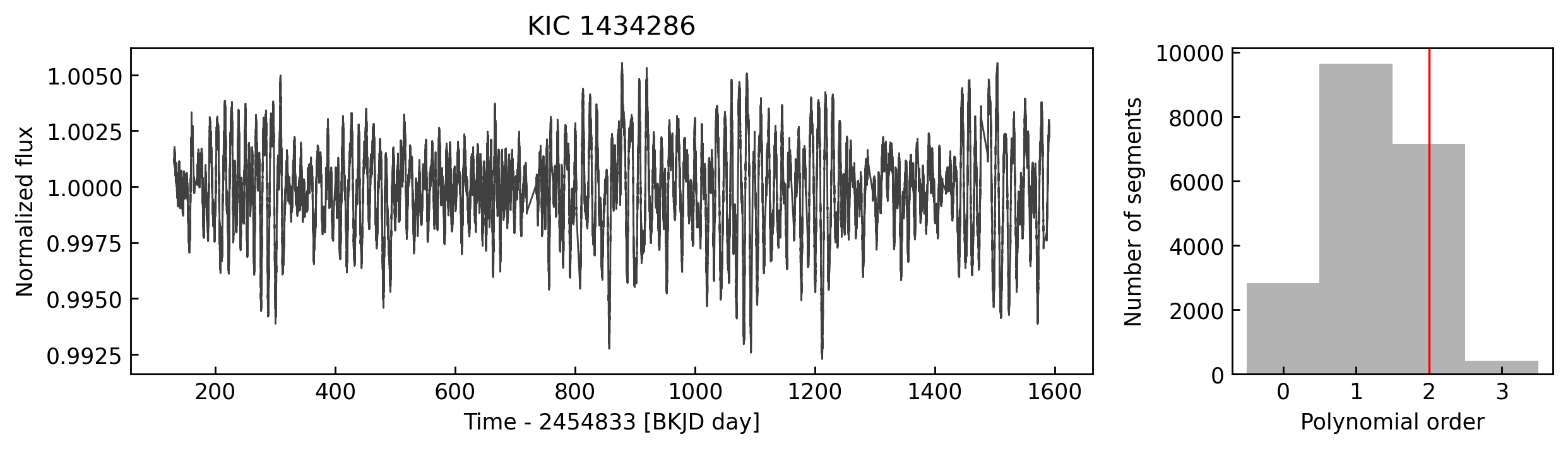}
\caption{Three Kepler light curves with different noise levels from non-transit samples and their frequency histograms of the best trend order for each segment. The red line represents the optimal trend order for the entire light curve. 
\label{fig3.2: noise model of three examples}}
\end{figure}

In the SPLS searching, a 4th-order polynomial for transits and an adaptive-order polynomial described in Section \ref{subsub: Model comparison of the stellar trend} for the background trends are chosen. Figure \ref{fig3.2: noise model of three examples} presents the results of the trend model comparison for three test light curves (before injection) with different noise levels. The right column displays the frequency histograms of the best trend order for each segment, with a window size equal to $3 \times d_{\rm inject}$, and the best global order at the 90th percentile. 

The data between quarters and within a single quarter has time gaps due to monthly data downlinks, safe modes, and loss of fine pointing. Our internal tests showed that these gaps can distort the SPLS periodogram and mimic the transit signal. Therefore, in addition to dividing the data into segments based on the window size, data points at both ends of larger gaps will also be separated and fitted independently. The large gap is quantified by a subjective threshold. In the comparison, we set the threshold to 4.5 times the observation cadence (29.4 min), meaning that the time interval within each fitted segment will not exceed this value during the SPLS search.

SDE serves as the statistical metric for comparing the periodograms of the BLS, TLS, and SPLS algorithms. Different algorithms use different quantities to compute the SDE. TLS employs the embedded normalized chi-squared, SPLS uses the normalized log-likelihood difference described in Section \ref{subsec: Statistic for periodogram}, and BLS also adopts the normalized log-likelihood difference which is derived by normalizing the ``likelihood" power output from the \texttt{BoxLeastSquares} in \texttt{Astropy}. This ``likelihood" power is an approximate log-likelihood difference between the transit box model and its baseline model which is calculated only using the data within the duration for computational efficiency. Although the definitions of SDE are different for different pipelines, they are equivalent in terms of showing the significance of a signal by scaling the likelihood in combination with a choice of threshold.

\subsubsection{Results}\label{subsubsection: results}

An effective tool for assessing the performance of different classifiers and determining threshold values is the Receiver Operating Characteristic (ROC) curve, which shows the relationship between the true positive rate (TPR) and the false positive rate (FPR) at every possible measure (here SDE) threshold. When calculating the ROC curve, the samples without injected planets are selected as actual negative samples, and the sample with injected transits is selected as an actual positive sample only if the period corresponding to the highest peak in the periodogram differs from the injected period (or its half or double) by no more than $2Pd_{\rm inject}/S$. This value is derived from Eq. (4) of \citealt{2014A&A...561A.138O} and $P$ refers to 0.5, 1, or 2 times $P_{\rm inject}$. We add a factor of 2 here to expand the period coverage in peak searching. TPR reflects the proportion of actual positive samples correctly identified by the classifier and FPR reflects the proportion of actual negative samples incorrectly recognized as positive at a given threshold. In the ROC curve, the diagonal line (TPR = FPR) represents random guessing with no classification ability. A classifier with better performance should have a higher TPR and a lower FPR, meaning that the ROC curve should be as close as possible to the top-left corner. Area Under the Curve (AUC), which ranges from 0 to 1, can also serve as a quantitative performance metric \citep{BRADLEY19971145}. A higher AUC value indicates that the classifier is more effective at distinguishing between positive and negative classes. An AUC of 1, 0.5, and less than 0.5 implies perfect, random, and incorrect predictions, respectively. Additionally, we use another supplementary metric: the TPR value at an FPR of 10 percent \citep{2024MNRAS.528.4053W}.

\begin{figure}[ht!]
\centering
\includegraphics[width=\linewidth]
{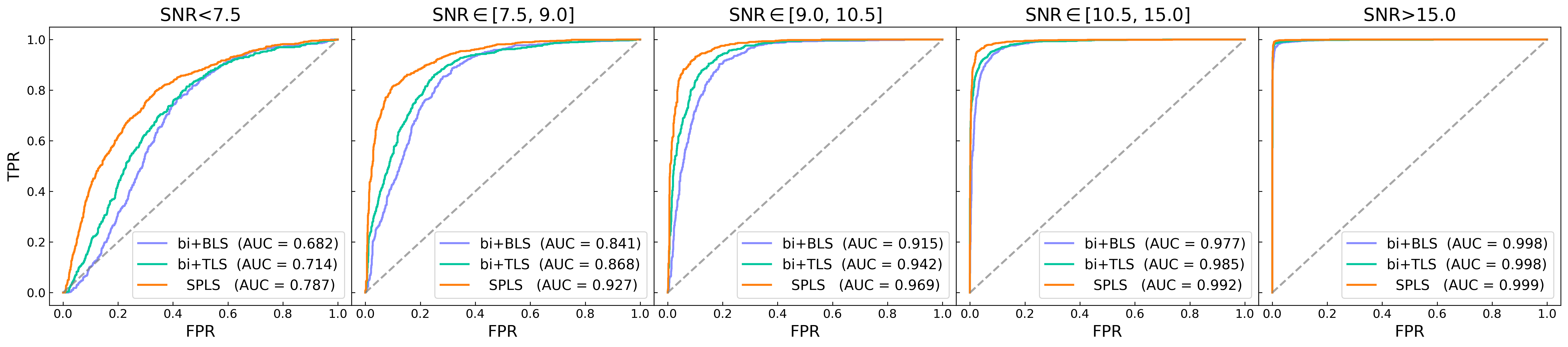}
\caption{ROC curves for SPLS (orange), biweight+BLS (purple), and biweight+TLS (green) in the SNR dimension. The gray dashed line represents the performance of a random classifier. For simplicity, ``biweight" is abbreviated as ``bi".
\label{fig: 3_3_1_ROC_SNR}}
\end{figure}

We investigated the detection sensitivity of three methods to transit signals in four dimensions, namely the SNR, planet radius, orbital period and the noise level of light curves. First, in the SNR dimension, the ROC curves in Fig. \ref{fig: 3_3_1_ROC_SNR} show the performance of the methods across five SNR bins, and the AUC and TPR at 10\% FPR in the top row of Fig. \ref{fig: 3_3_1_ROC_statis_SNR_R} are presented as a function of these bins. From these figures, SPLS consistently outperforms the other methods, especially in detecting low-SNR signals. For example, in the SNR ranges of $<7.5$ and [7.5, 9.0], SPLS improves the AUC by 0.073 and 0.059 over biweight+TLS, respectively. Also at a 10\% FPR, SPLS achieves TPRs that are 22.6\% and 26.7\% higher than those of biweight+TLS. When the SNR exceeds 15, the performance of all methods becomes comparable.

\begin{figure}[ht!]
\centering
\includegraphics[width=0.85\linewidth]
{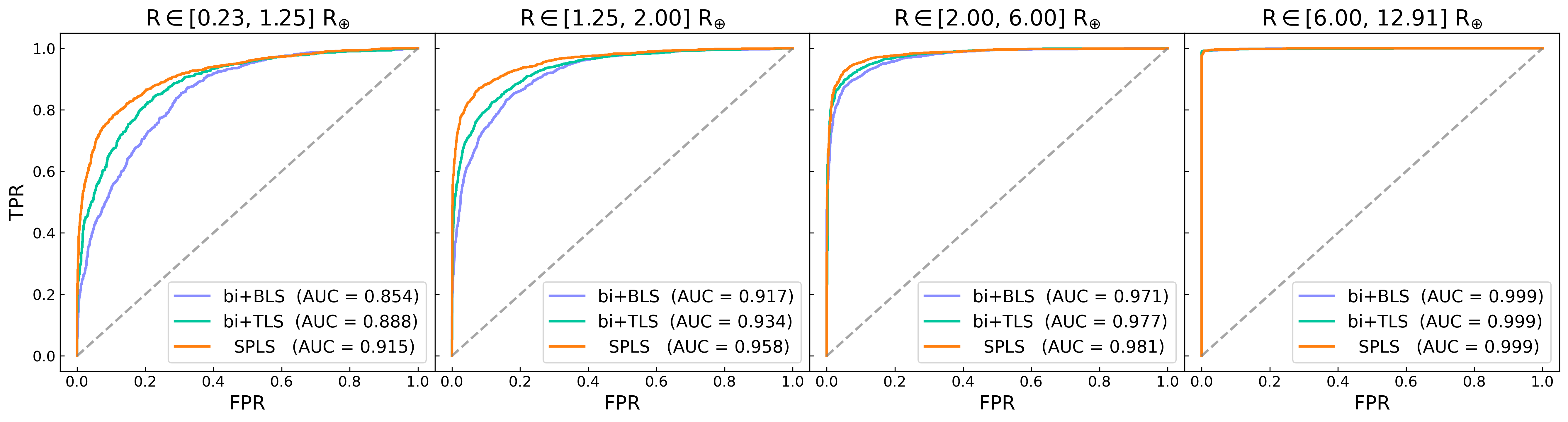}
\caption{ROC curves for SPLS, biweight+BLS, and biweight+TLS in the planetary radius dimension. The four radius bins correspond to Earth-sized, super-Earth-sized, Neptune-sized, and Jupiter-sized planets.
\label{fig: 3_3_1_ROC_R}}
\end{figure}

\begin{figure}[ht!]
\centering
\includegraphics[width=0.7\linewidth]
{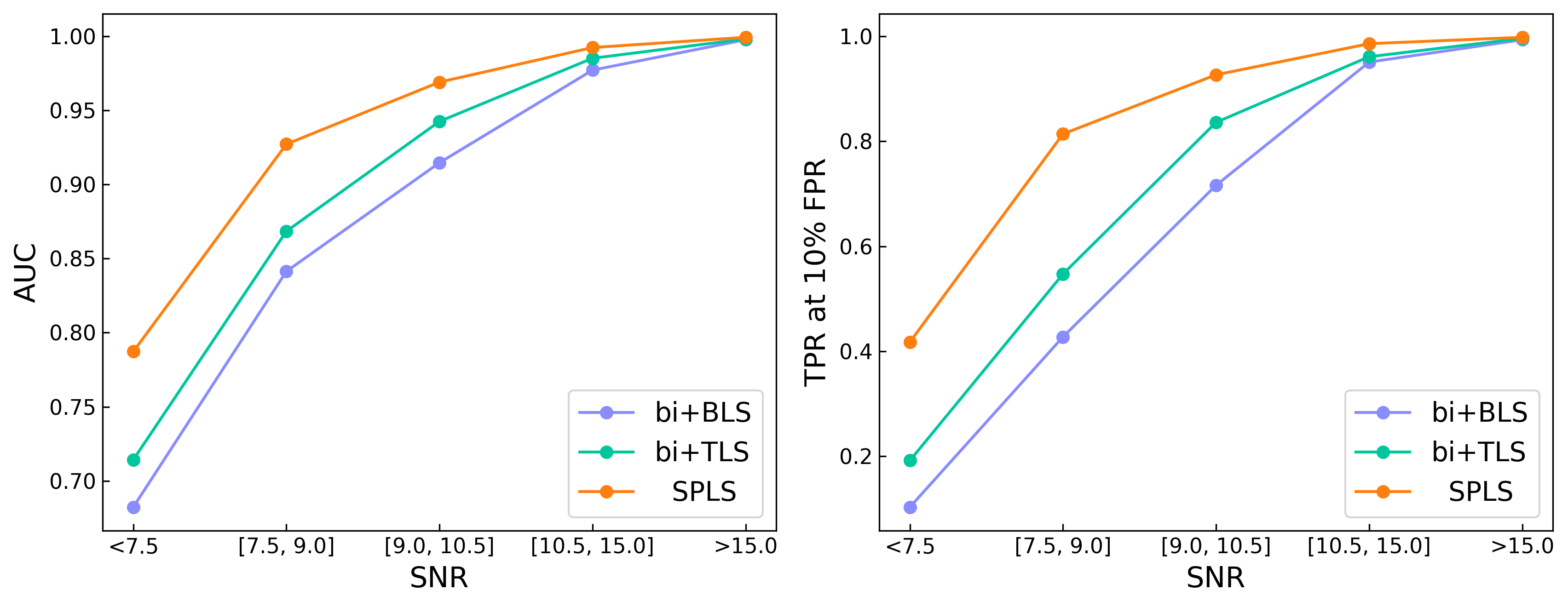}
\includegraphics[width=0.7\linewidth]
{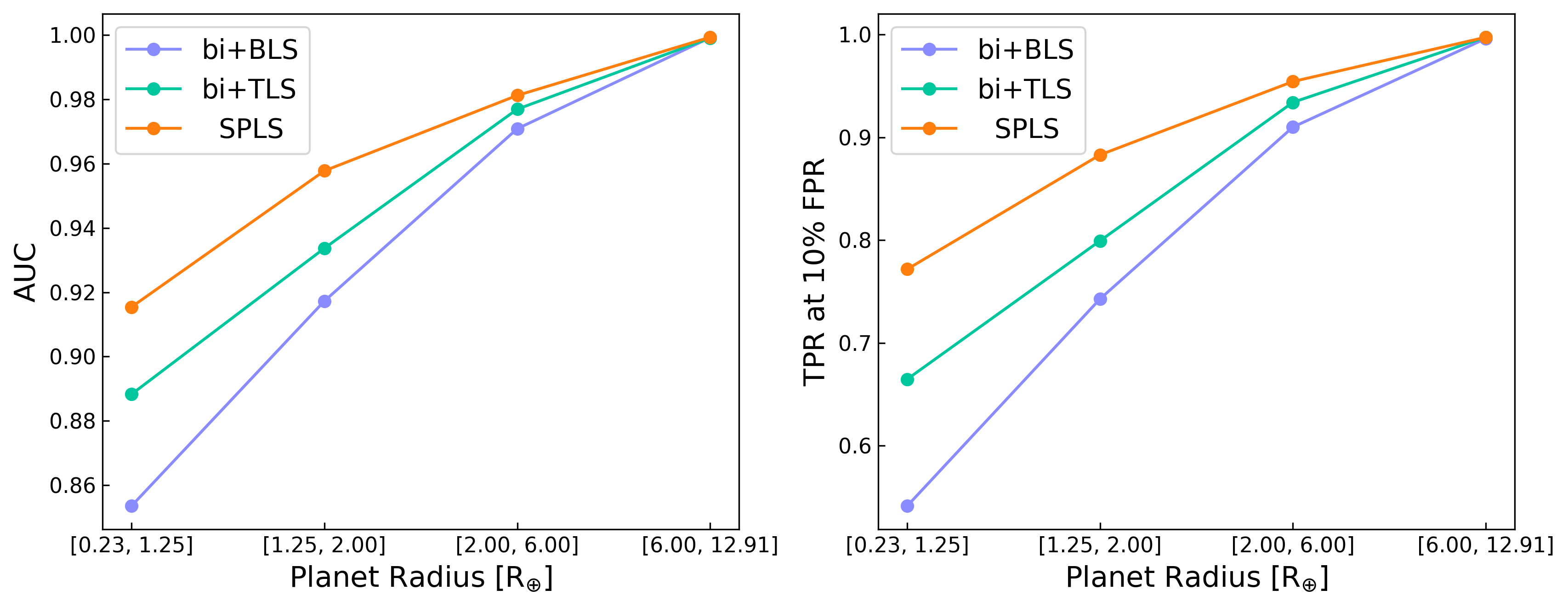}
\caption{AUC (left) and TPR at 10\% FPR (right) as functions of SNR bins (top row) and planet radius bins (bottom row) for SPLS, biweight+BLS, and biweight+TLS.
\label{fig: 3_3_1_ROC_statis_SNR_R}}
\end{figure}

Although SNR is the most direct factor affecting detection performance, the planet radius, which reflects the intrinsic physical size, provides an intuitive perspective. Therefore, we also include the ROC curves across different planet radius bins as a supplementary comparison (Fig. \ref{fig: 3_3_1_ROC_R} and the bottom row of Fig. \ref{fig: 3_3_1_ROC_statis_SNR_R}). The bins correspond to Earth-sized ($R_{\rm p}$ $<$ 1.25 R$_{\oplus}$), super-Earth-sized (1.25 R$_{\oplus}$ $\leq$ $R_{\rm p}$ $<$ 2 R$_{\oplus}$), Neptune-sized (2 R$_{\oplus}$ $\leq$ $R_{\rm p}$ $<$ 6 R$_{\oplus}$), and Jupiter-sized ($R_{\rm p}$ $\geq$ 6 R$_{\oplus}$) planets. Our method achieves the best performance for Earth-sized, super-Earth-sized, and Neptune-sized planets, with the highest AUC and TPR at 10\% FPR compared to the other methods. All methods perform similarly for Jupiter-sized planets.

\begin{figure}[ht!]
\centering
\includegraphics[width=0.7\linewidth]
{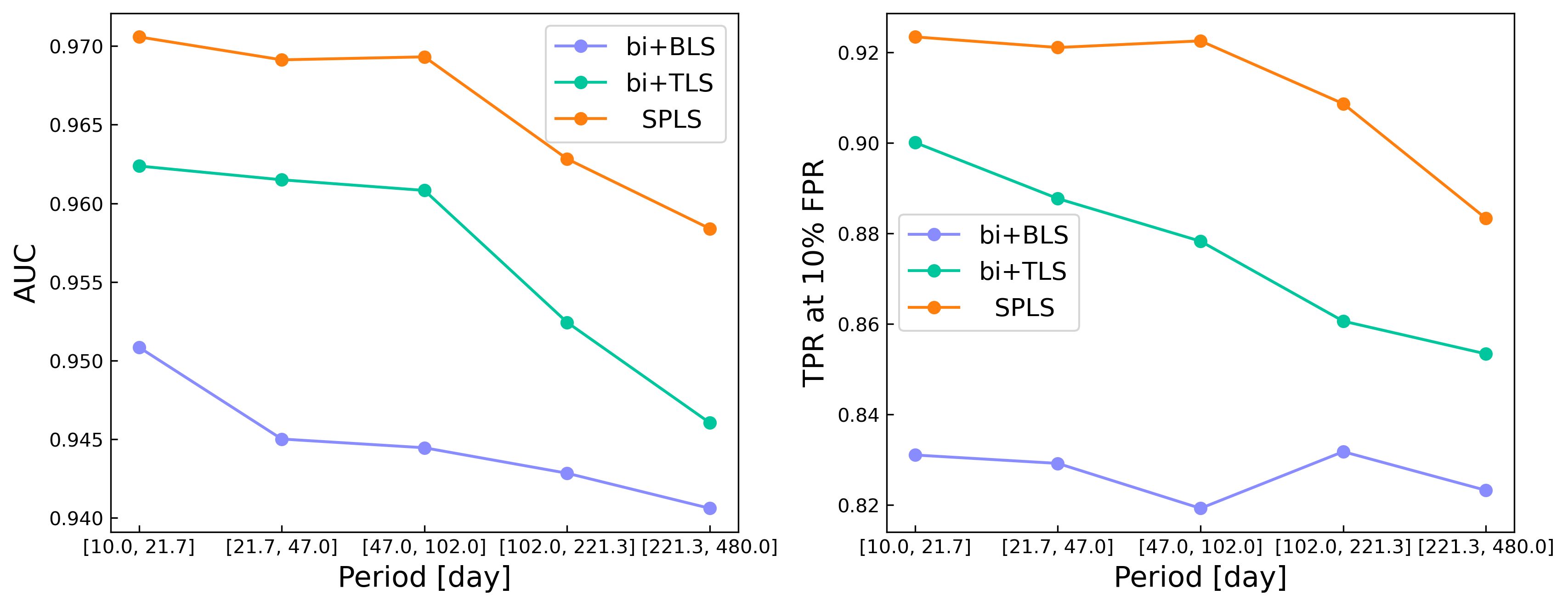}
\includegraphics[width=0.7\linewidth]
{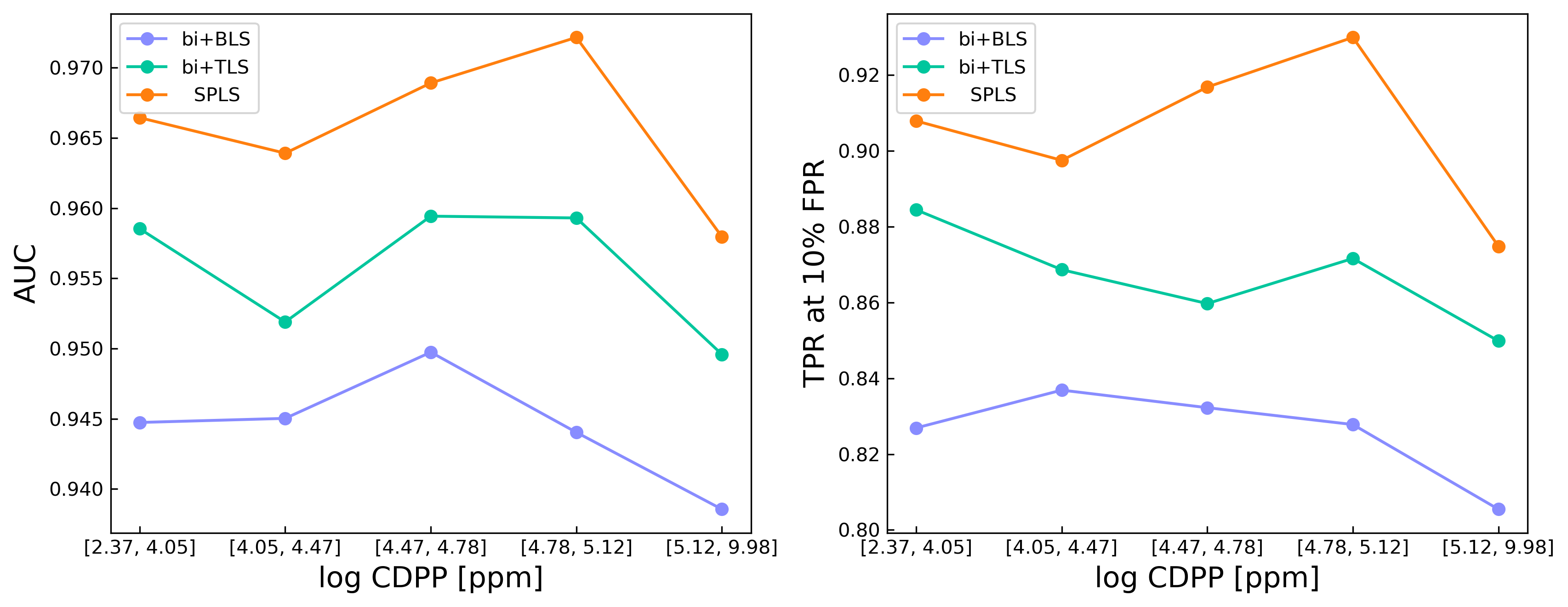}
\caption{AUC (left) and TPR at 10\% FPR (right) as functions of period bins (top row) and noise-level bins (bottom row) for SPLS, biweight+BLS, and biweight+TLS.
\label{fig: 3_3_1_ROC_statis_P_CDPP}}
\end{figure}

In addition, we examined the results in the period and noise-level dimensions (see Fig. \ref{fig: 3_3_1_ROC_P_CDPP} in the Appendix \ref{app: ROCs} for the ROC curves). The orbital periods were divided evenly into five bins in logarithmic space, and the noise levels were divided into five bins based on the quintiles of $\log\ {\rm CDPP}$. As shown in Fig. \ref{fig: 3_3_1_ROC_statis_P_CDPP}, our method achieves the highest AUC and TPR at 10\% FPR across all period and noise-level groups, indicating that SPLS performs best among three methods.

While ROC curves provide valuable insight into the classification capability of detection methods, they are inherently limited to a univariate analysis of true versus false signals. The ROC analysis above assumes that the injected orbital period, or its direct alias, produces the strongest signal for each actual positive sample, and then treats the periodogram power threshold as the varying single variable. In practice, however, planetary detection typically requires two simultaneous conditions: the dominant periodogram peak must occur near the injected period (or its direct alias), and its power must exceed a predefined threshold. Therefore, the ROC curve comparison alone lacks a comparative assessment of different methods' ability to recover the true period.

\begin{figure}[ht!]
\centering
\includegraphics[width=0.5\linewidth]
{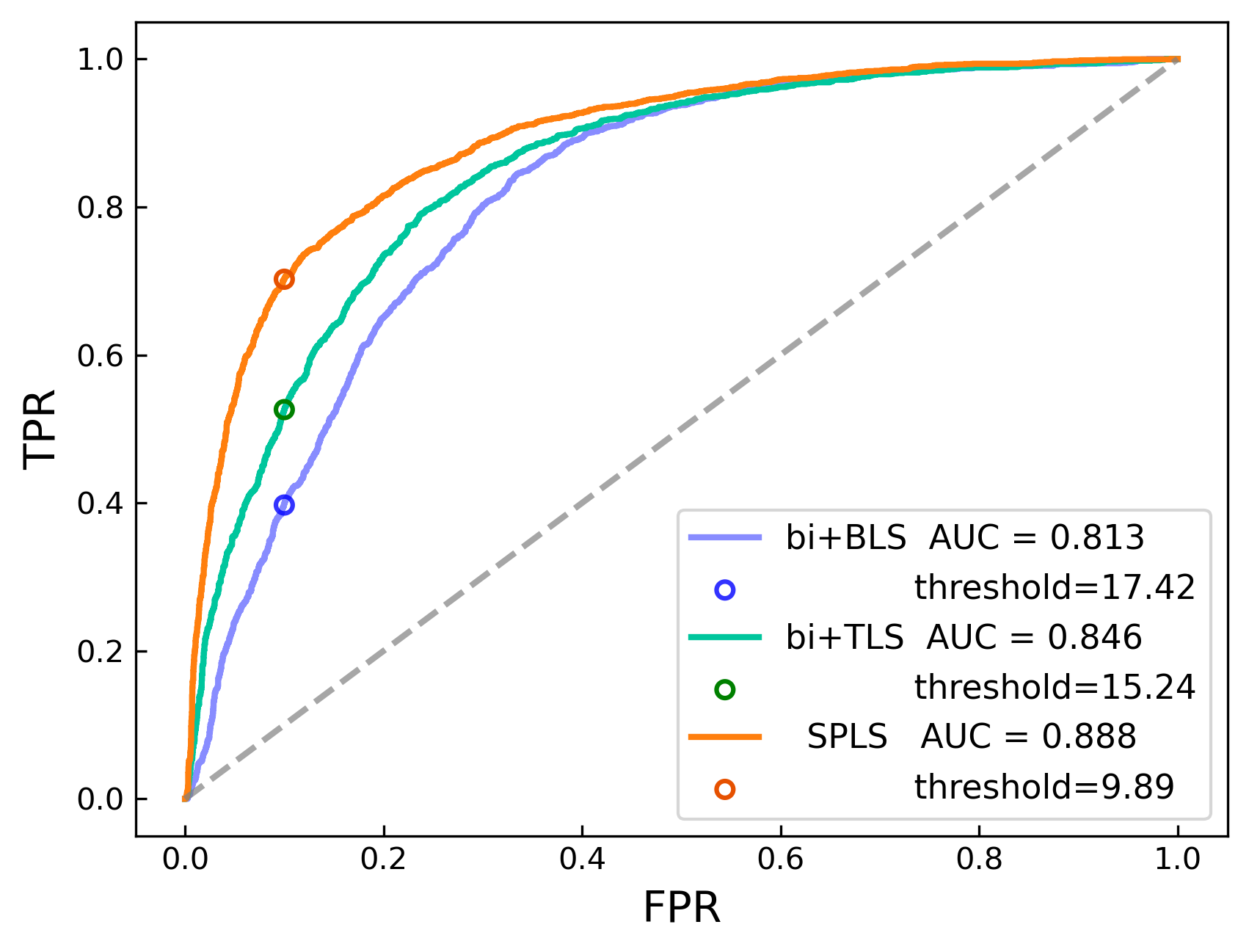}
\caption{ROC curves for SPLS, biweight+BLS, and biweight+TLS using injected samples with SNR below 10 and their corresponding non-injected samples. Circles represent the thresholds at 10\% FPR.
\label{fig: 3_3_2_ROC_thresholds}}
\end{figure}

To comprehensively evaluate our method, in addition to the ROC curve analysis, we also included another performance comparison that examines the recovery rate of the 10,000 injected light curves for different methods. We define the recovery rate as the proportion of injected samples where the recovered period corresponding to the highest peak in the periodogram deviates from the injected period (or its half or double) by no more than $2Pd_{\rm inject}/S$, with the peak's SDE above a threshold. The false recovery rate is the proportion of samples where the highest peak occurs at an incorrect period but still exceeds the SDE threshold. For the three methods, we applied different thresholds at 10\% FPR of their ROC curves, as shown in Fig. \ref{fig: 3_3_2_ROC_thresholds}. The ROC curves here for threshold selection were generated using injected samples with SNR below 10 and their corresponding non-injected samples, which leads to lower thresholds and allows for a better comparison of the methods' sensitivity in recovering weak signals.

\begin{figure}[ht!]
\centering
\includegraphics[width=\linewidth]
{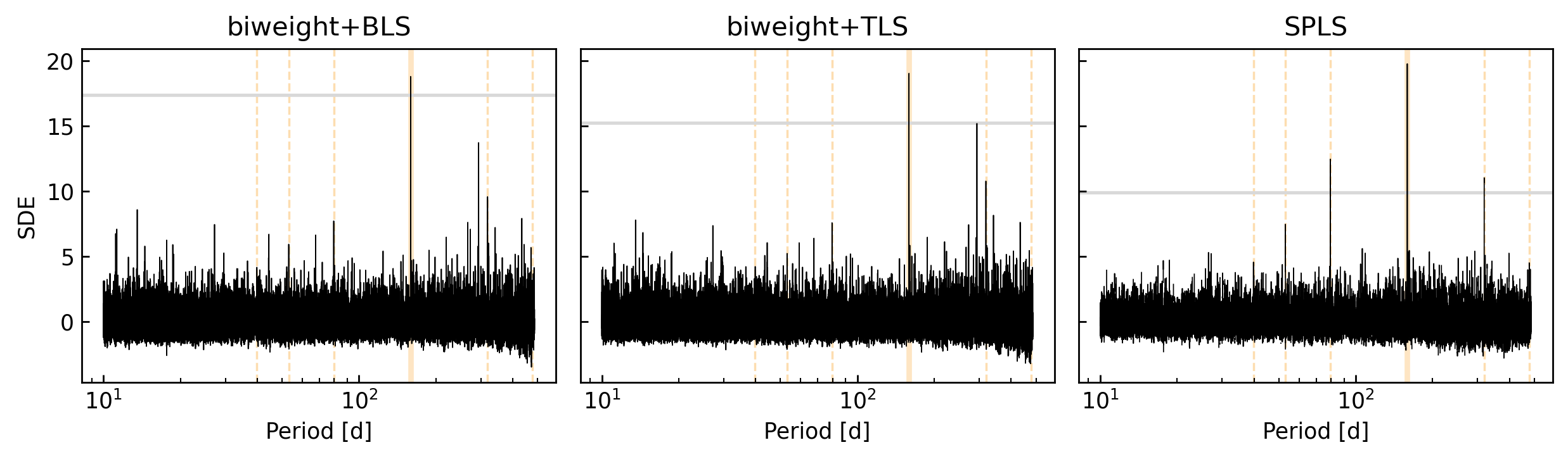}
\includegraphics[width=\linewidth]
{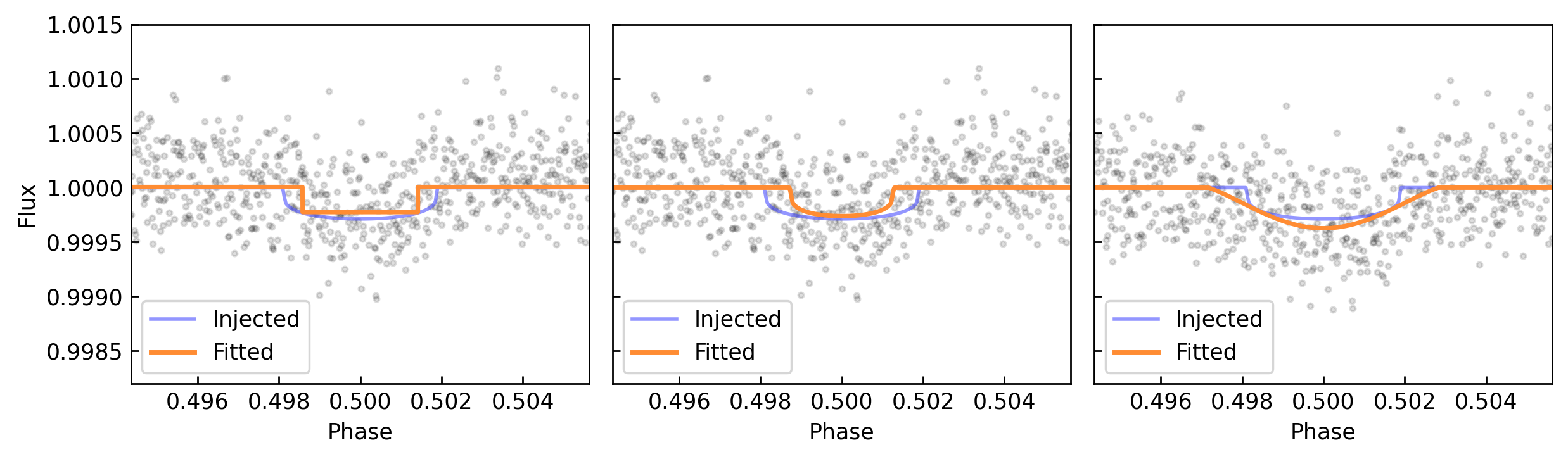}
\caption{Periodograms and fitted phase curves of KIC 10841614 with an injected transiting planet, having a SNR of 9.02 and an orbital period of 159.23216 days, obtained by three methods. The specific injection process is described in Section \ref{subsubsection: Test data and preprocessing}. In the top row, the horizontal gray lines show the thresholds determined from Fig. \ref{fig: 3_3_2_ROC_thresholds}. The vertical yellow lines represent the period at the highest peak (solid) and its aliases (dashed). In the bottom row, the orange line denotes the fitted phase curve, while the purple line denotes the injected signal. To visualize the fitting results in a manner similar to BLS and TLS, the data points and fitted curve in the SPLS phase plot are adjusted by deliberately subtracting the fitted trend component of the periodic model.  
\label{fig: 3_3_KIC10841614}}
\end{figure}

As an illustrative example from our injection–recovery test, we present the periodograms and fitted phase curves of KIC 10841614, whose injected planet (SNR = 9.02, period = 159.23216 day) is recovered by all three methods (Fig. \ref{fig: 3_3_KIC10841614}). In the periodograms, the horizontal gray lines mark the SDE thresholds at 10\% FPR as discussed above. SPLS yields the lowest threshold and the most significant peak relative to it, highlighting its advantage in detecting transits and reducing false positives. In the phase-folded light curves, the transits fitted by BLS and TLS exhibit shorter durations than the injected signal. Given that their search grids cover the true duration, this bias is likely caused by detrending distortions on weak signals. By contrast, the SPLS fit shows a slightly longer duration and a less consistent shape compared to the injected signal, owing to the flexibility of its double polynomial model, which introduces degeneracy among the duration parameter and the signal and background polynomial coefficients. The typical procedure for transit detection begins with identifying the signal using the SPLS periodogram, followed by refining parameter estimates through MCMC. Therefore, precise recovery of the transit shape is not critical during the SPLS search stage. It is also worth noting that the SPLS curve shown is not derived from a fit to the detrended data. Instead, both the signal and background are modeled simultaneously. The plotted data and fitted curves here have just had the fitted background component removed for a clearer comparison with the biweight+BLS/TLS results.

\begin{figure}[ht!]
\centering
\includegraphics[width=0.8\linewidth]
{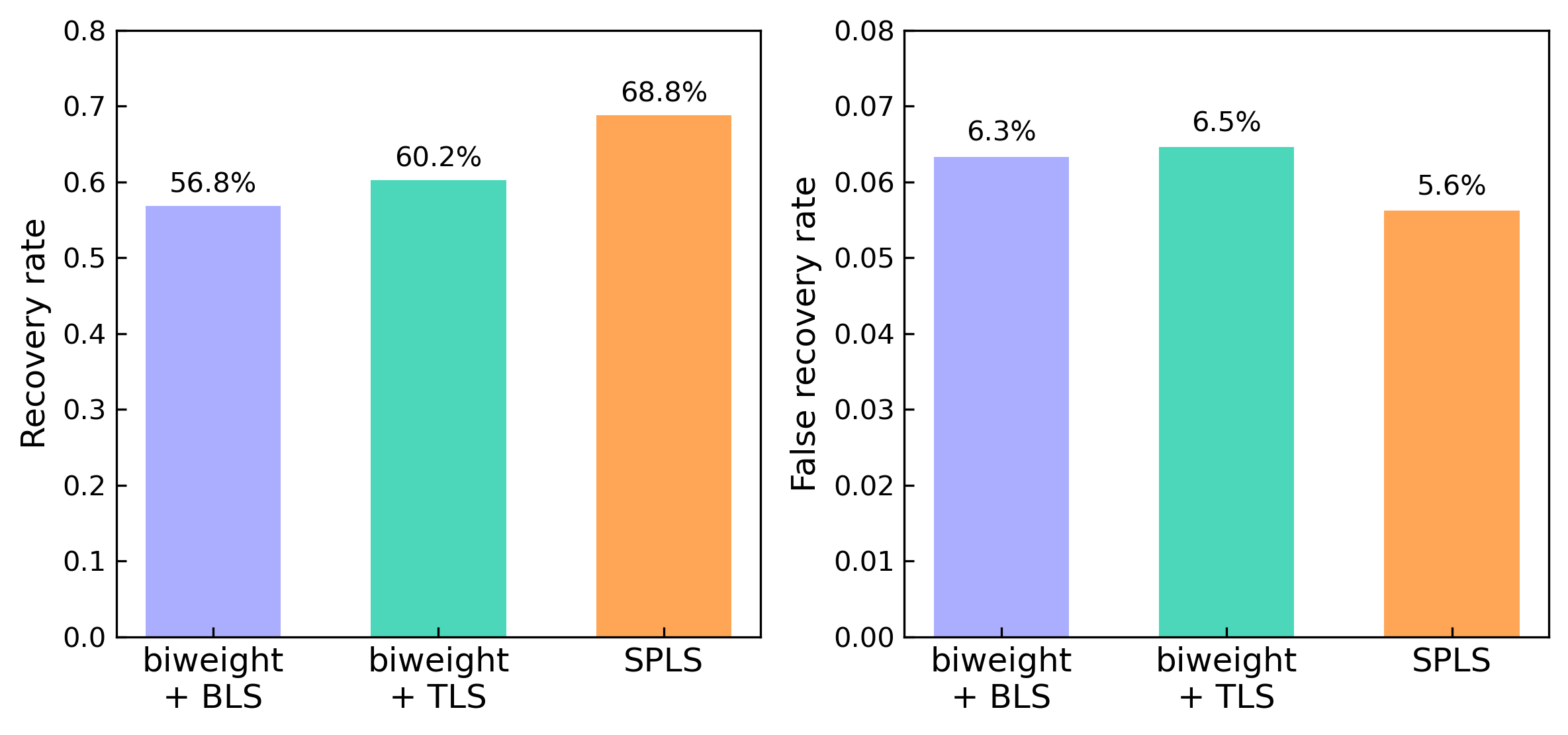}
\caption{The recovery rate and false recovery rate of 10,000 samples with injected transits for three methods.
\label{fig: 3_3_2_all}}
\end{figure}

\begin{figure}[ht!]
\centering
\includegraphics[width=\linewidth]
{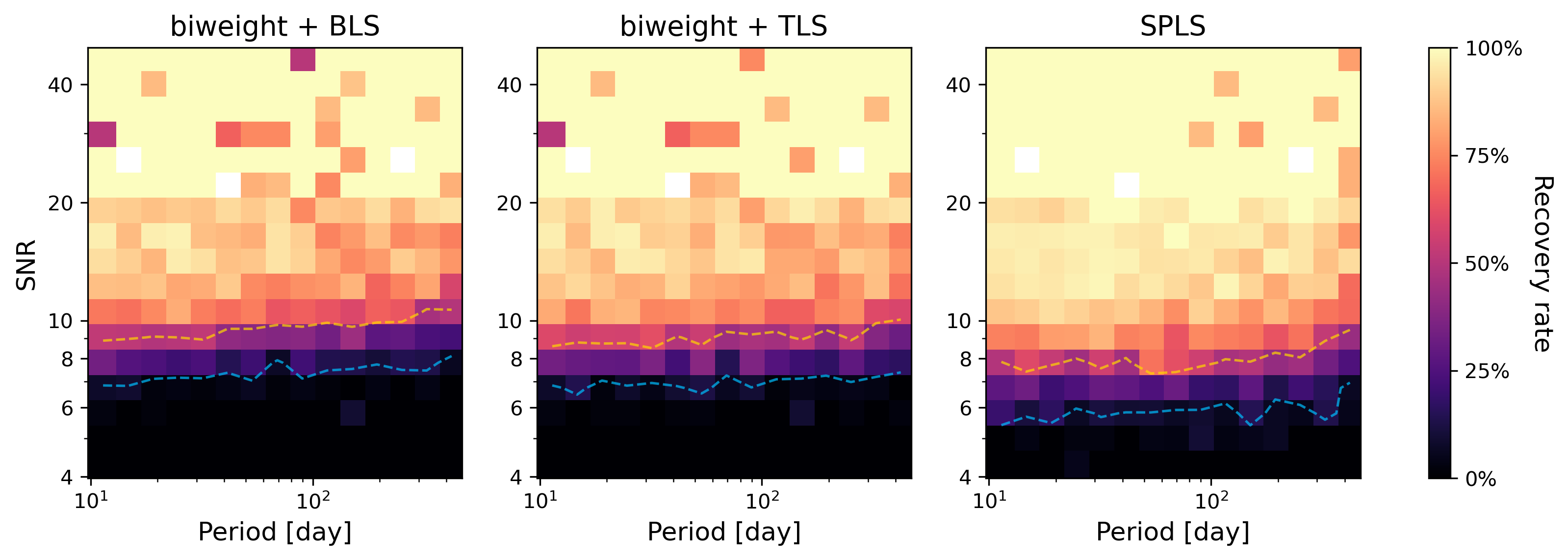}
\caption{The recovery rate of samples with injected transits in SNR and orbital period space for the three methods. The SNR is log-uniformly grouped into 17 bins, and the orbital period is divided into 15 bins. The injected SNR range is broad, but for clearer visualization and comparison, only the range from 4 to 50 is shown. Blank areas indicate bins with no samples. The dashed blue and gold contour lines represent the 10\% and 50\% recovery rates, respectively. 
\label{fig: 3_3_2_P_SNR}}
\end{figure}

\begin{figure}[ht!]
\centering
\includegraphics[width=\linewidth]
{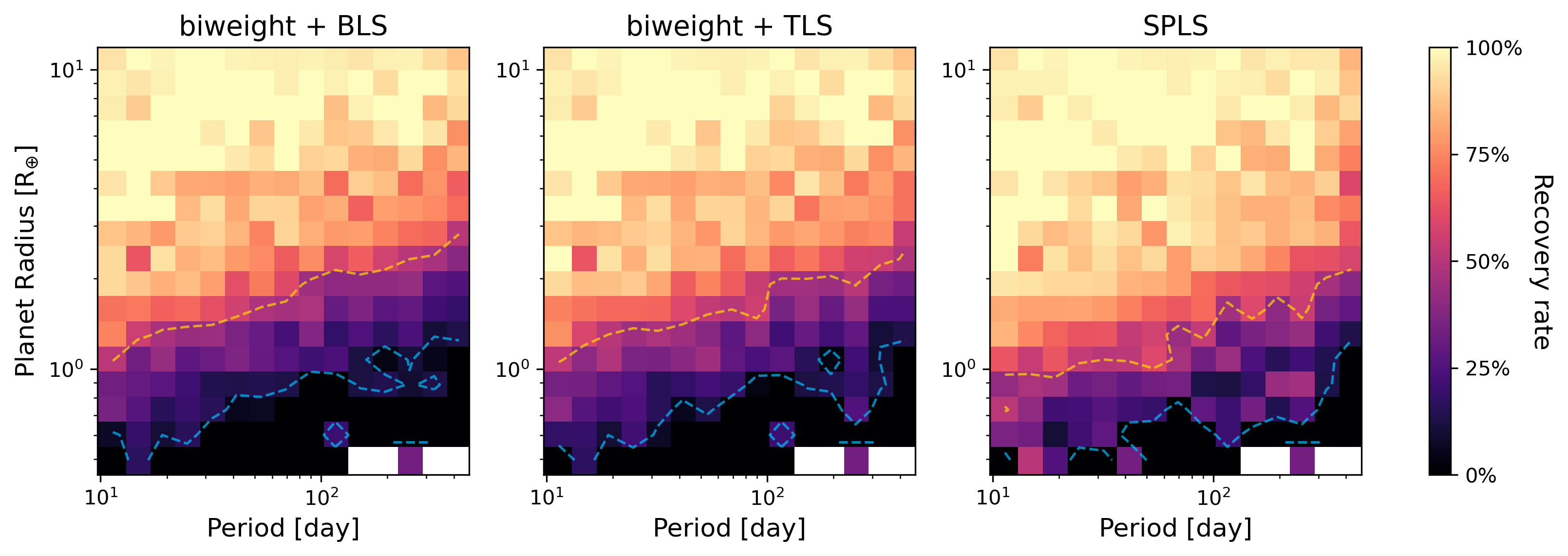}
\caption{The recovery rate of samples with injected transits in planet radius and orbital period space for the three methods. The planet radius is log-uniformly grouped into 17 bins, and the orbital period is divided into 15 bins. Blank areas indicate bins with no samples. The dashed blue and gold contour lines represent the 10\% and 50\% recovery rates, respectively.
\label{fig: 3_3_2_P_Rp}}
\end{figure}

The recovery rate and false recovery rate for the three methods are visualized in Fig. \ref{fig: 3_3_2_all}. SPLS exhibits the highest recovery rate and the lowest false recovery rate, demonstrating that our method continues to outperform the others when considering both period recovery accuracy and the given SDE thresholds. We also plotted the results in the injected SNR and orbital period space which is grouped into a 17 $\times$ 15 grid of logarithmically uniform bins (Fig. \ref{fig: 3_3_2_P_SNR}). Since the false recovery rates of the three methods show little variation in the two-dimensional space, only the recovery rate distribution is displayed. The dashed blue and gold contour lines represent the 10\% and 50\% recovery rates, respectively. The results show that the advantage of SPLS holds across almost all ranges in the space. It is particularly evident in its enhanced sensitivity to low-SNR transit signals. For example, SPLS exhibits a downward shift of the 10\% recovery rate to lower values (5-6) compared to other methods. Similarly, the results in the injected planet radius and orbital period space (Fig. \ref{fig: 3_3_2_P_Rp}) also show that our method lowers the detection limit for small-radius planets, which makes the application of our algorithm for detecting Earth twins seem promising.

We performed a simple comparison of the computational times of three pipelines. Based on the settings of the three methods described in Section \ref{subsubsection: settings of three methods}, the CPU times for each target were $\sim$0.5 minutes for biweight+BLS, $\sim$50.8 minutes for TLS, and $\sim$193.2 minutes for SPLS, all of which were run on a single CPU of an Apple M2 chip device. The large-scale tests in this section were conducted on the Intel Xeon ICX Platinum 8358 processors of the Siyuan-1 cluster\footnote{\url{https://docs.hpc.sjtu.edu.cn/en/index.html}} supported by the Center for High Performance Computing at Shanghai Jiao Tong University. It is evident that the better detection sensitivity of our algorithm also results in a higher computational cost. Therefore, we recommend using our method as a supplementary approach. For deep transit signals, traditional detrending-detection methods are prioritized, while our method focuses primarily on detecting weak signals. Future work may involve utilizing GPU acceleration or other methods to further speed up the computations.

\subsection{Application to Kepler single-planet systems}\label{subsection: application to Kepler Single-Planet Systems}

\begin{figure}[ht!]
\centering
\includegraphics[width=0.5\linewidth]
{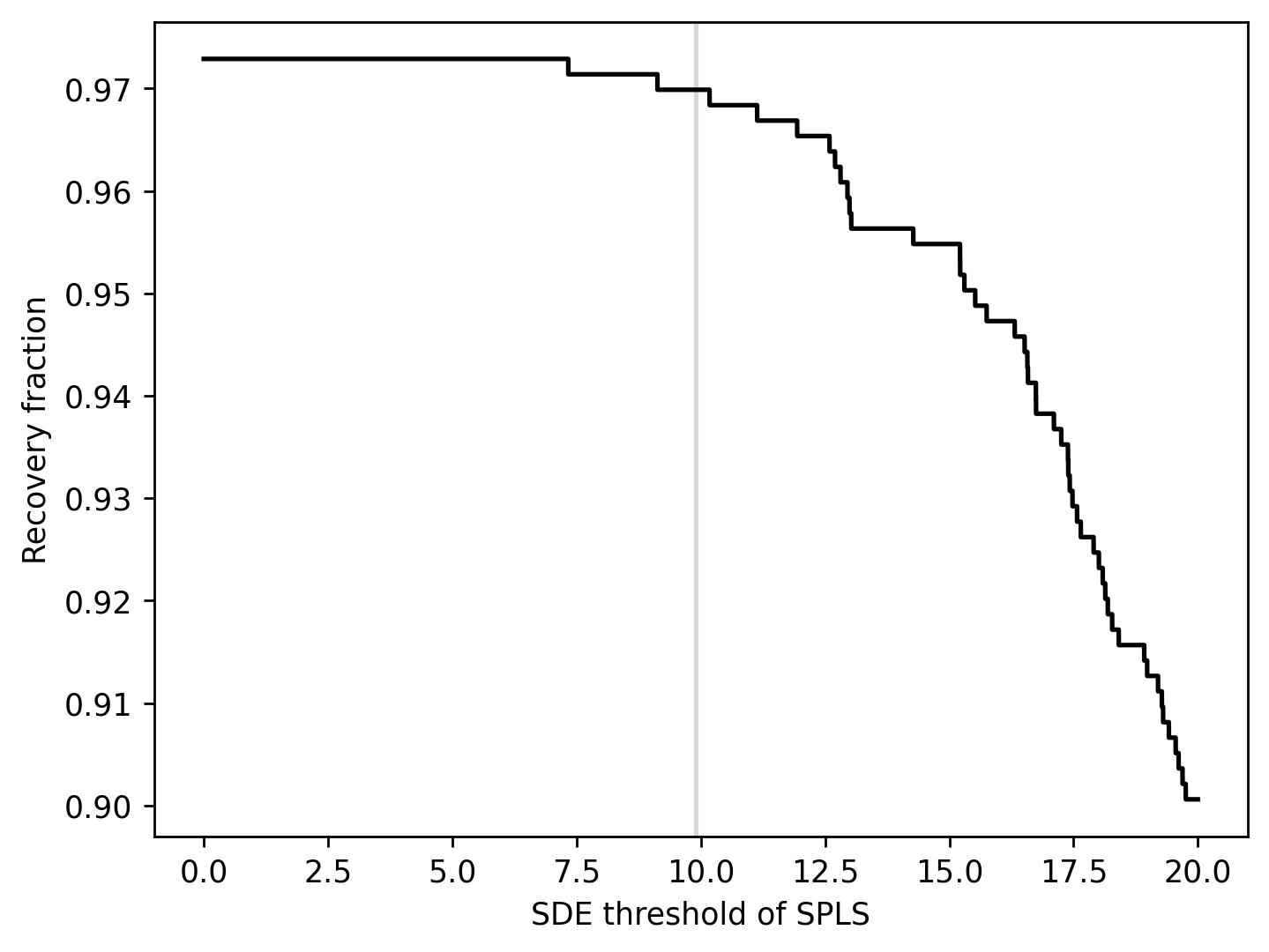}
\caption{Recovery fraction of SPLS as a function of the SDE threshold. The grey vertical line marks the threshold derived from Fig. \ref{fig: 3_3_2_ROC_thresholds}, corresponding to a value of 9.89.
\label{fig: 3_4_real}}
\end{figure}

Our algorithm was also applied to the Kepler confirmed single-planet systems to assess its ability to recover known signals. We selected targets from 9,564 dispositioned KOIs in the NASA Exoplanet Archive through two selection criteria. First, we retained only systems with a single confirmed planet and no additional candidates. Second, we required the targets to have a transit duration longer than the minimum duration sampled by SPLS by default, an orbital period greater than 10 days, and at least two observed transits. A total of 664 targets satisfied these criteria. We adopted the long-cadence PDC light curves from Quarters 1–17, and the preprocessing followed the same procedure as described in the last paragraph of Section \ref{subsubsection: Test data and preprocessing}. The SPLS search parameters were also set as described in Section \ref{subsubsection: settings of three methods}, except that the maximum search period was extended to the full time span of the data. Based on these settings, the recovery fraction as a function of the SDE threshold for SPLS is shown in Fig. \ref{fig: 3_4_real}. A 97.0\% recovery fraction is achieved at a threshold of 9.89 (the grey vertical line), indicating the effectiveness of the approach. The few unrecovered cases are mainly caused by factors such as outlier handling, unused transits near custom time gap edges, and the selection of background trend polynomials. These issues are discussed in detail in the Discussion section. 

\section{Discussion}\label{sec: discussion}
\subsection{Advantages of SPLS in detecting weak long-period signals}

The analysis of ROC curves and the recovery performance of injected transit samples in the injection–recovery test demonstrate that SPLS maintains improved detection capability across a wide range of SNR, planet radius, orbital period, and light-curve noise levels, compared to the commonly used detrending-detection methods. Notably, the enhanced sensitivity to low-SNR, long-period ($>$ 10 days) transits is particularly valuable for future searches for Earth twins. Additionally, the examination of long-period signals using our method is also an aspect that previous simultaneous fitting algorithms have not explored.

\begin{figure}[ht!]
\centering
\includegraphics[width=\linewidth]
{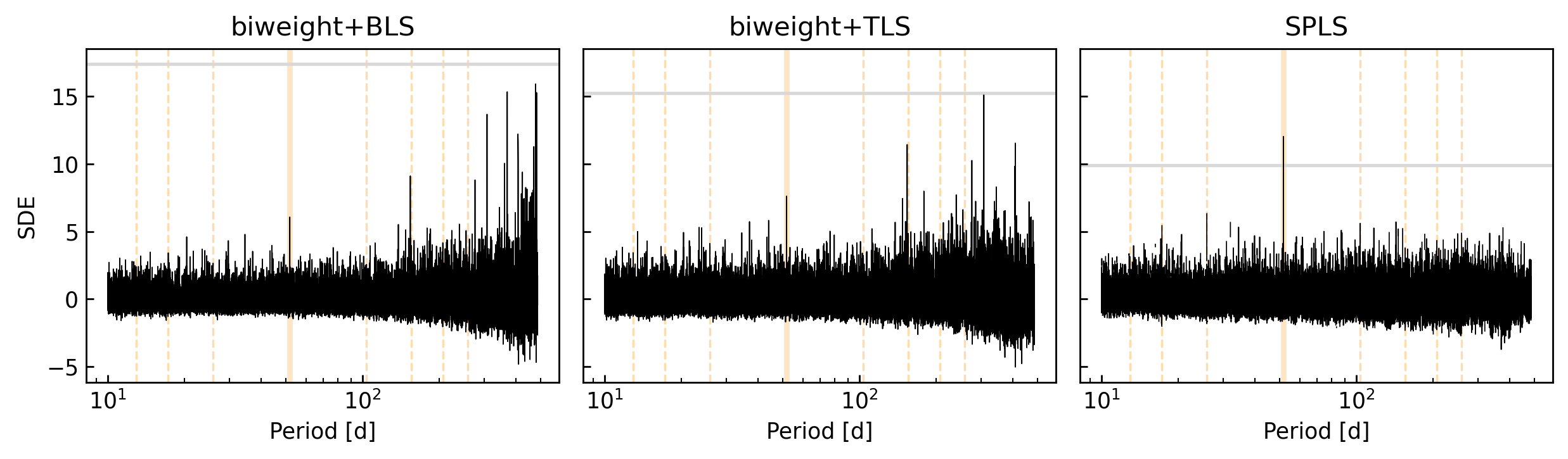}
\caption{Three periodograms of the Kepler long-cadence light curve of KIC 8565138 which contains 24 transits originating from an injected planet with an SNR of 6.650 (a radius of 1.284 R$_{\oplus}$) and a 51.787-day orbital period. The gray lines represent the thresholds that are the same as those of Fig. \ref{fig: 3_3_KIC10841614}. The vertical yellow lines represent the period and its aliases of the injected planet.
\label{fig: 4_1_KIC8565138}}
\end{figure}

The advantage remains even when some transits near time gap edges are missed by SPLS. In order to prevent numerical instability in the SPLS polynomial model, sufficient numbers of data points must be present within the trial transit duration as well as on both sides outside the duration to solve the polynomial coefficients. Therefore, transit data near predefined gaps is not used by our method, which means that SPLS utilizes fewer transits than the number of injected transits at the ideal parameters of period, epoch and duration, while this issue does not arise with the other two methods in the comparison of Section \ref{section: performance} \footnote{Actually whether this issue occurs in detrending-detection methods depends on the detrending method used.}. We counted the number of samples in which SPLS successfully recovered the transit signals, but biweight+BLS and biweight+TLS did not, totaling 1,004 samples. Among these, 702 samples exhibit the characteristic that SPLS uses fewer transits at the optimal nonlinear parameters, suggesting that our method maintains superior detection performance. Figure \ref{fig: 4_1_KIC8565138} shows an example that the planet with 24 transits was injected into KIC 8565138 and SPLS only used 22 transits, but the SPLS periodogram shows the highest sensitivity to the signal and suppressed strength of false signals at large periods. The negative influence of this characteristic is minor, since only about 0.4\% of targets, mostly high-SNR signals, fail to be detected due to a decrease in signal strength when SPLS is applied to Kepler confirmed planets.

The improved detection capability is attributed to our algorithm design. Our combined background trends and signal modeling enhances sensitivity to each transit, making the overall periodic signal more prominent without the risk of signal attenuation caused by detrending methods. Additionally, the segmented local fitting allows SPLS to focus on noise within the window rather than longer-term correlated noise, likely helping to reduce false positives or alarms in the long-period range. Although the exclusion of transit near predefined time gaps loses partial transit information, it also excludes false alarms to a certain extent.

Currently, in blind searches, the maximum sampled duration needs to be selected by the user based on their detection goals. We recommend setting the window size to at least twice $d_{\rm max}$ to ensure that the signal portion does not overly dominate the model. In our tests, we deliberately chose a window size of three times the injected duration, the same as the window size for biweight. However, while three times injected duration is ideal for biweight, it may not be the best choice for our algorithm. If a window size more suited to SPLS were selected, the performance of SPLS might improve further.

\subsection{Optimization of trend model} 

The optimal polynomial order for the background trend component is determined by model comparison based on the Bayes factor for each segment. We statistically analyze the distribution of optimal orders for all the uniformly sampled segments, and use the order at the 90th percentile as the final global order for the light curve. But is it appropriate to use the order at the 90th percentile? The final selected trend order should ideally describe the variability of the segments at any position as accurately as possible. However, since the variation of background noise is stochastic, different trial transit segments prefer different trend orders as shown in the distribution (Fig. \ref{fig: N_orders}). Selecting a relatively small order may result in inadequate trend modeling in some segments and generating false positives, while selecting a larger order may also lead to over-fitting in some segments. Users have the flexibility to select their preferred order of the trend component. The 90th percentile is a subjective choice in our tests. The results in Section \ref{section: performance} show that SPLS performs effectively in capturing background trends and reducing false positives in most cases, suggesting that the 90th percentile is generally sufficient.

\begin{figure}[ht!]
\centering
\includegraphics[width=\linewidth]
{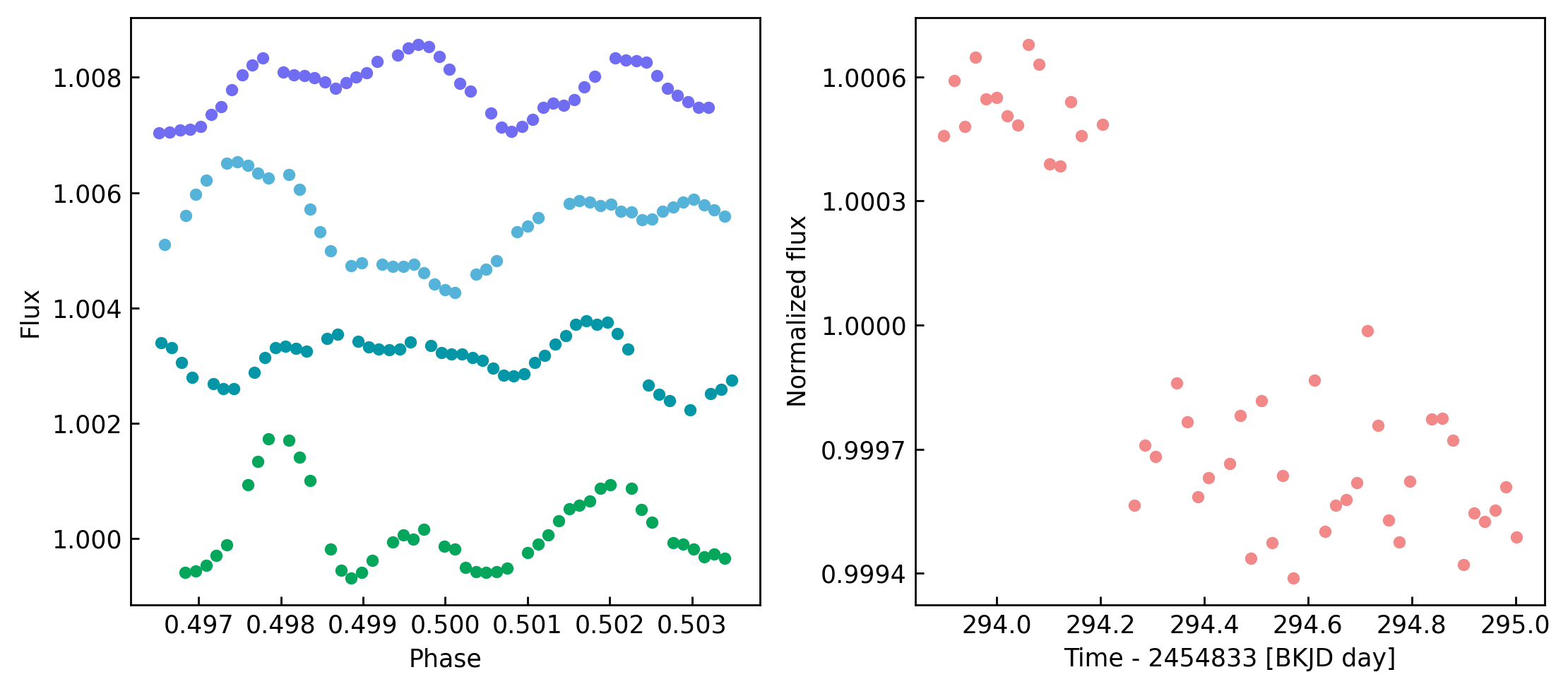}
\caption{Four segments of KIC 2424934's Kepler light curve folded at $P$ = 162.07 d, $t_{m0}$ = 2455187.196 d (left), and a segment of KIC 2299081's Kepler light curve centered on 2455127.447 d (right). The window sizes for both panels are 1.134 days. The normalized flux of the four segments is vertically shifted for clarity in the left panel.
\label{fig3_3+test3: SPLSno}}
\end{figure}

In our injection–recovery test, our algorithm performs well in detecting transits in light curves with different noise levels. However, this result is based on stellar samples from which most evolved stars have been excluded. In our internal tests, we also attempted to apply SPLS to light curves with higher noise, but the method performs poorly when the background variability within a fitted window exhibits high-frequency fluctuations. The left panel of Fig. \ref{fig3_3+test3: SPLSno} shows high-frequency fluctuations in several segments without transits for a set of nonlinear parameters. The polynomial order for the trends of this target was determined to be the third order. However, this order is insufficient to accurately describe the trends, impacting the detection of injected transits. Regardless of whether we increase or decrease the polynomial order of the trend component, the results remain poor. A higher background order risks removing the signal together with the trends, whereas a lower order fails to adequately capture the noise pattern. Reducing the SPLS window can simplify noise in certain light curves, but this also restricts our ability to detect longer transit durations. Hence, we consider that limitation can not be addressed by the simple double-polynomial model. In conventional detrending-detection pipelines, partial short-memory variations remain after detrending is also a common issue (e.g., \citealt{2019AJ....158...57C}; \citealt{2024AJ....167..202M}). For weak signal detection in the light curves whose background variability timescales are much shorter than the window size of SPLS, complex noise models such as ARMA (AutoRegressive Moving Average), ARIMA (AutoRegressive Integrated Moving Average), and Gaussian processes may offer better suitability. However, incorporating these models with more nonlinear parameters into the joint signal and noise fitting framework significantly introduces a higher computational burden.

\subsection{Dealing with gaps in light curves}

We defined a threshold for time gaps to ensure that each fitted window contained no large gaps. The threshold value warrants further consideration, because a large value may lead to polynomial overfitting and introduce spurious signals, while a small value may split a full transit and reduce the available signal information. Moreover, the choice of the value is also related to the data preprocessing procedure, since preprocessing can disrupt data continuity. For example, in the recovery of the two transits of the long-period Kepler-421 b, the preprocessing step caused a time gap in one of the transits due to the removal of a portion of data flagged as bad quality. As a result, this transit was split, which affected the signal detection. Overall, our default threshold value (4.5 times the observation cadence) is deemed acceptable, given SPLS's superior performance in the injection-recovery test.

The presence of flux gaps (including outliers) within a fitted window can also produce false positives. The right panel of Fig. \ref{fig3_3+test3: SPLSno} presents a case, even though no time gaps exceed the preset threshold. SPLS easily misidentifies this flux gap as a deep transit. We found that among the confirmed planets not detected by SPLS, eight were missed due to this reason, as they can be successfully recovered after handling the gap flux. The simple handling we used for the flux gaps here is as follows: A flux gap is considered significant if the difference in flux between adjacent time points exceeds five times the Median Absolute Deviation (MAD)-based standard deviation. The data is then split to ensure that no significant flux gaps remain within each fitted segment. Notably, the handling is simple, and its main limitation lies in the difficulty of distinguishing true transits from false signals caused by flux gaps, leading to the removal of deep transits as well. Therefore, this processing should only be applied when searching for shallow signals. Future research into more robust flux gap handling techniques is also worth further investigation.

\subsection{Scope of Applicability} 

Our algorithm is well-suited for approximately uniform data, especially data from space-based telescopes. As discussed in the preceding section, our algorithm offers superior detection capability for low-SNR, long-period transiting planets from the light curves where the background variability timescale is larger than the window size of SPLS, which is also the range where we recommend using SPLS. While SPLS is not suitable for detecting short-period signals within long-baseline data, which is mainly caused by two reasons. On the one hand, the number of sampled periods and mid-transit times depends on the time span and the set minimum period, as shown in Fig. \ref{fig: dP}. For example, searching for periods shorter than 10 days in the Kepler four-year long-cadence data can generate a $P-t_m$ grid containing up to $10^{13}$ combinations, resulting in extremely high and impractical computational load. One the other hand, the linear search of our three-step approximation gives a cadence-dependent limit to the minimum duration that requires each trial transit in the periodic signal to have a sufficient number of data points for individual fitting. Then the corresponding window size and trial period should not be small. Therefore, for detecting weak short-period signals in long-baseline datasets, we recommend algorithms such as GPFC \citep{2024MNRAS.528.4053W} and APRS \citep{2019AJ....158...57C}, which can cumulative transit information by phase folding and have demonstrated strong capabilities in detecting them. 

\subsection{Future work} 

There are several aspects we can explore in future work. First, SPLS can be applied to the light curves of Kepler, K2, TESS, and future missions such as Earth 2.0 and PLATO to discover Earth twins. Second, cotrending technique is used to remove systematic trends. For example, Cotrending Basis Vectors is a commonly used tool in the PDC unit of Kepler and TESS pipelines. Similar to detrending, cotrending can also damage transit signals, reducing the possibility of subsequent detection of weak signals. We intend to incorporate systematic noise modeling into the existing periodic model to further improve the detection of weak transiting signals. Third, the efficient parallel computing power of Graphics Processing Unit (GPU) is increasingly being used in planetary exploration (\citealt{2024MNRAS.528.4053W}; \citealt{2024AJ....167..284G}; \citealt{2025MNRAS.tmp..496S}), and we also plan to develop a GPU version of SPLS for faster processing. Fourth, our tests focused only on the detection of single-planet systems. For multi-planet system detection, one possible approach is to remove the strongest planetary signal from the 2D grid of the linear search and to conduct a second periodic search for the potential second planet. Additionally, the linear search also makes the detection of the single transit feasible. The details will be explored in the future. In addition, another direction is exploring how to extend SPLS to apply to unevenly sampled datasets.

\section{Conclusion}\label{sec: conclusion}

We present a novel algorithm, SPLS, designed to detect weak long-period transiting exoplanets from light curves with background trends by simultaneously modeling both planetary transits and trends, thereby facilitating the search for Earth twins. A segmented double polynomial is introduced to suppress the effect of long-term correlated noise and enhance sensitivity to individual transits. Prior to the search, the optimal polynomial order for modeling the trend component in each light curve is determined using Bayes factor-based model comparison.

We evaluate the performance of SPLS against conventional detrending-detection methods through an injection-recovery test, in which synthetic transits with varying SNRs (planet radii) and orbital periods are added to real non-transiting Kepler light curves. The detection sensitivity of SPLS consistently outperforms other methods across a broad range of SNRs, planet radii, orbital periods, and light-curve noise levels. For example, SPLS achieves the highest AUC and TPR at a 10\% FPR for detecting Earth-sized, super-Earth-sized, and Neptune-sized planets, as well as for long periods ranging from 10 to 480 days in 4-year spanning light curves.

Additionally, we examine the recovery capability of SPLS in 10,000 injected samples. With the thresholds determined from the ROC analysis, SPLS exhibits the highest recovery rate (8.6\% higher than biweight+TLS) and the lowest false recovery rate. SPLS also pushes the SNR sensitivity threshold at 10\% TPR down to 5-6, further underscoring its advantage in detecting weaker signals. In conclusion, these results emphasize the potential of SPLS in the search for Earth twins.

Although the algorithm is accelerated through a three-step approximation and the sampling of nonlinear parameters is optimized for compatibility with SPLS, its improved sensitivity still comes at the cost of increased computational demand compared to standard detrending-detection approaches. In the future, we aim to further accelerate the algorithm through GPU-based implementation. Overall, its remarkable performance makes it a valuable tool for detecting more weak long-period planets in both current and future space-based missions.


\begin{acknowledgments}

We thank the referees for their insightful comments and valuable suggestions that greatly improved our paper. This work is supported by the National Key R\&D Program of China, No. 2024YFC2207700 and No. 2024YFA1611801, by the National Natural Science Foundation of China (NSFC) under Grant No. 12473066, by the Shanghai Jiao Tong University 2030 Initiative, by SJTU-Warwick Joint Seed Fund 2024/25-Round 5, and by the China-Chile Joint Research Fund (CCJRF No. 2205). CCJRF is provided by Chinese Academy of Sciences South America Center for Astronomy (CASSACA) and established by National Astronomical Observatories, Chinese Academy of Sciences (NAOC) and Chilean Astronomy Society (SOCHIAS) to support China-Chile collaborations in astronomy. The computations in this paper were run on the Siyuan-1 cluster supported by the Center for High Performance Computing at Shanghai Jiao Tong University. This research has made use of the NASA Exoplanet Archive, which is operated by the California Institute of Technology, under contract with the National Aeronautics and Space Administration under the Exoplanet Exploration Program. This paper includes data collected by the Kepler mission. Funding for the Kepler mission is provided by the NASA Science Mission Directorate. 

\end{acknowledgments}

%



\software{numpy \citep{harris2020array}, numba \citep{siu_kwan_lam_2025_15182388}, matplotlib \citep{Hunter:2007}, lightkurve \citep{2018ascl.soft12013L}, Astropy \citep{astropy:2013, astropy:2018, astropy:2022}, pandas (\citealt{mckinney-proc-scipy-2010}; \citealt{reback2020pandas}), wotan \citep{2019AJ....158..143H}, TLS \citep{2019A&A...623A..39H}, SPLS \citep{shuyue_zheng_2025_15411397}}

\facility{Kepler}

   



\appendix
\section{Additional ROC curves}\label{app: ROCs}

\begin{figure}[ht!]
\centering
\includegraphics[width=\linewidth]
{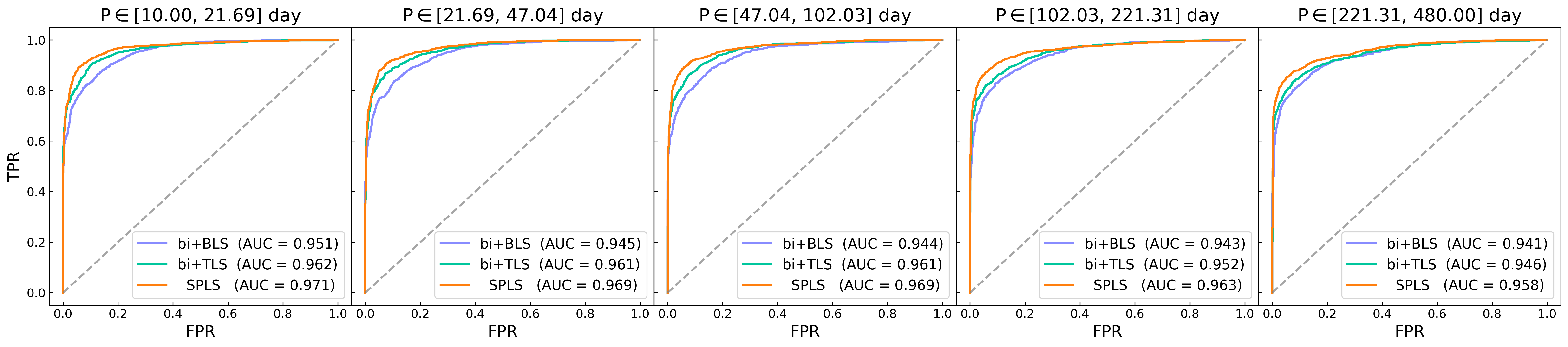}
\includegraphics[width=\linewidth]
{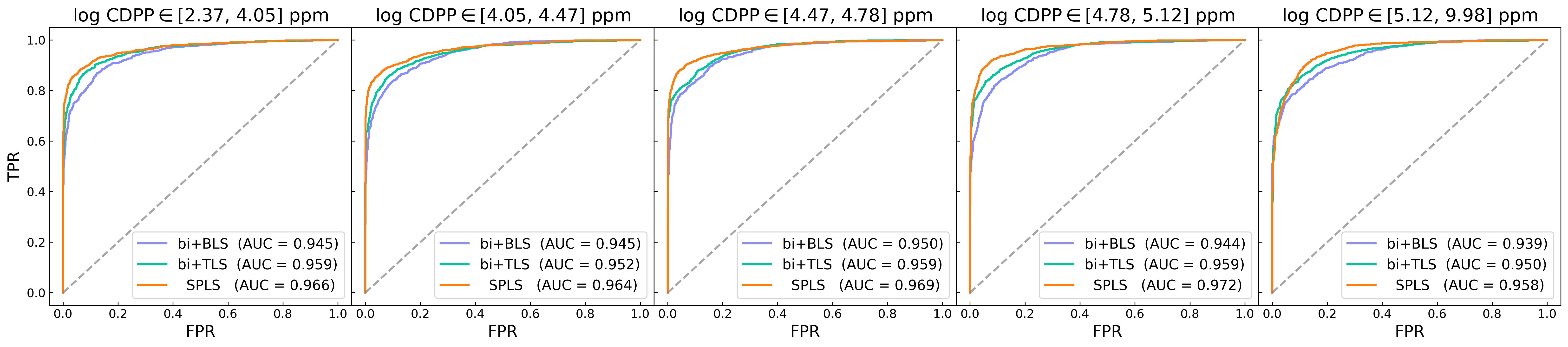}
\caption{ROC curves for SPLS, biweight+BLS, and biweight+TLS in the period dimension (top row) and CDPP dimension (bottom row).
\label{fig: 3_3_1_ROC_P_CDPP}}
\end{figure}

\bibliography{sample631}{}
\bibliographystyle{aasjournal}



\end{document}